\documentclass[]{minimal-journal}
\usepackage{amsmath}
\usepackage{amsfonts}
\usepackage{bbm}
\usepackage{graphicx}
\usepackage{hyperref}

\usepackage[nopatch]{microtype}
\usepackage{float} 
\usepackage{needspace} 
\usepackage{booktabs}
\usepackage{tabularx}
\usepackage{ragged2e}
\graphicspath{{figures/}}

\title{A Unified Bayesian Model for Voter Turnout Estimation: Combining Surveys, Aggregate Data, and Selection Correction}

\author{M. Niitsoo}
\affiliation{Liberal Citizen Foundation SALK, Estonia}
\alsoaffiliation{Institute of Computer Science, University of Tartu, Tartu}

\author{R. Rebane}
\affiliation{School of Economics and Business Administration, University of Tartu, Tartu}

\author{T. Jüristo}
\affiliation{Liberal Citizen Foundation SALK, Estonia}
\alsoaffiliation{Department of Stage Arts, Estonian Academy of Music and Theatre, Estonia}
\email[T. Jüristo]{tarmo@salk.ee}


\begin{document}

\begin{abstract}
Accurate small-area estimation of voter turnout for demographic subgroups is crucial for political analysis but methodologically challenging. Survey data suffer from over-reporting, non-representativeness, and non-ignorable non-response, while ecological inference (EI) from aggregate data is vulnerable to the ecological fallacy. We propose a Bayesian hierarchical framework that jointly integrates individual-level survey data, official aggregate turnout margins, and census cell counts. The framework comprises three models: a multilevel ecological inference (EI) model that extends MRP-style structure to aggregate data, a Poll-and-Margin (PM) model that jointly fits survey and margins in one coherent likelihood, and a Full Selection (FS) model that adds a Heckman-style selection correction under a random-contact assumption and uses informative priors for identification. In simulations with controlled selection mechanisms based on empirical census demographics, FS substantially improves over established benchmarks, particularly under strong selection bias or heavy censoring, but is sensitive to unmodeled selection noise. In an application to the 2023 Estonian parliamentary election, PM performs similarly to the current state of the art on available validation margins, with FS showing a modest improvement that depends on the informative priors.
\end{abstract}

\section{Introduction}
Understanding voter turnout across fine-grained demographic subgroups is essential for political scientists, policymakers, and campaign strategists \citep{leighley_who_2013}. As Lijphart \citep{lijphart1997unequal} has argued, the central democratic concern is not merely the level of turnout but rather who participates and who abstains. When participation is systematically lower among certain groups—particularly those with lower incomes, less education, or from marginalized communities—the electorate becomes unrepresentative and policy outcomes may favor the preferences of higher-propensity voters. Beyond questions of representation, accurate estimates of differential turnout inform electoral administration decisions, enable campaigns to allocate mobilization resources efficiently, and help policymakers understand which communities face barriers to participation \citep{leighley_who_2013}. Detecting and measuring these patterns requires estimation at the intersection of multiple demographic characteristics: understanding turnout among "low-income voters" is insufficient if patterns differ substantially between low-income college graduates and low-income non-graduates, or vary by race, age, and geography.

The value of granular differential turnout estimation extends beyond any single application. Electoral coalitions are increasingly fragmented along multiple intersecting dimensions—education, race, age, geography, and income interact in complex ways that cannot be captured by examining marginal effects of individual characteristics \citep{doherty_changing_2024}. A method that can flexibly estimate turnout for arbitrarily defined subgroups enables researchers to test competing theories about turnout patterns, identify emerging electoral trends, and detect whether established patterns are stable or shifting. Moreover, many practical applications require estimation at geographic scales smaller than those for which high-quality survey data are available: evaluating whether a state-level policy change affected turnout differently across counties, for instance, demands methods that can generate reliable small-area estimates from limited data.

In principle, two main data sources exist for this determination: survey data and official aggregate turnout statistics. Survey-based approaches, such as Multilevel Regression and Poststratification (MRP) \citep{park_bayesian_2004}, leverage rich individual-level covariates. MRP estimates subgroup outcomes by combining a multilevel modeling stage with population aggregation via poststratification, using census data to weight estimates according to true population proportions. This approach partially pools information across groups, stabilizing estimates for sparse cells. However, surveys are plagued by declining response rates \citep{brick_williams_2013} and non-ignorable non-response bias, where respondents systematically differ from non-respondents \citep{groves_peytcheva_2008, groves_role_2004}. In political surveys, for instance, individuals more interested in politics are both more likely to respond and more likely to vote, creating selection bias.

Conversely, methods relying on turnout statistics, known as ecological inference (EI), infer individual behavior from aggregate data, thus avoiding \textit{survey non-response selection} bias but introducing the risk of ecological fallacy \citep{robinson1950ecological}. The fundamental challenge of EI is that many different configurations of individual-level behavior are consistent with observed aggregate totals. While methods like the Bayesian approach in \citet{king_solution_1997} incorporate logical bounds, inferences remain sensitive to assumptions about the distribution of behavior across groups \citep{cho_iff_1998, freedman_ecological_1999}. Moreover, aggregate data alone cannot recover detailed demographic interactions without strong modeling assumptions.

Each approach has complementary strengths and weaknesses. Surveys provide individual-level covariate patterns but biased samples; aggregate data provides unbiased population totals but no individual links. Intuitively, combining both should yield better estimates. Joint treatments are nevertheless rare, and the two closest precedents come from different fields.

In epidemiology, \citet{jackson_improving_2006} combine aggregate (ecological) disease counts with a supplementary sample of individual-level records in a single hierarchical Bayesian model in which both likelihoods share the same regression coefficients, showing that even a modest individual-level sample can substantially reduce the indeterminacy of aggregate-only inference; the approach was later generalized as hierarchical related regression \citep{jackson_hierarchical_2008}. This shared-coefficient joint likelihood is the same core mechanism our joint models employ. Their setting, however, assumes the individual-level sample is representative of the population (missingness is ignorable) and involves a small number of covariates, whereas turnout estimation must confront non-ignorable survey non-response, systematic over-reporting, and a fine-grained demographic cross-classification.

In political science, \citet{ghitza_deep_2013} proposed a postprocessing calibration step to MRP estimates to match known aggregates, a very significant step but an \textit{ad hoc} procedure that does not formally model the data-generating process end-to-end. Instead, their method applies a per-region scalar shift to estimated cell probabilities so that aggregated totals match official counts, thus implicitly attributing the mismatch to regional idiosyncrasies. This correction is also done in a frequentist way (no priors) and separately from the main model fitting, so it does not preserve statistical properties like uncertainty intervals.

Neither precedent addresses the non-ignorable non-response introduced above: while poststratification corrects imbalance on observed demographics, it implicitly assumes missingness is ignorable within cells. Selection models targeting exactly this problem have a long tradition in econometrics \citep{heckman1978, heckman1979}, but their use in survey settings is often constrained by the need for exclusion restrictions or specialized auxiliary data \citep{bailey_countering_2025, sciarini-overreporting-heckman}.

This paper proposes a systematic, model-based integration of survey data, aggregate turnout margins, and census information within a unified Bayesian framework. We introduce three models: a multilevel ecological inference (EI) model based on MRP, a Poll-and-Margin (PM) model that jointly fits survey and aggregate data, and a Full Selection (FS) model that additionally corrects for non-ignorable survey non-response using a Heckman-style selection mechanism under a comparatively weak assumption on survey design (random contact). We evaluate our models through an extensive simulation study, benchmark them against established methods, and apply them to data from the 2023 Estonian parliamentary election.

\section{Existing approaches}
Small-area turnout estimation aims to infer cell-level participation probabilities for fine-grained demographic groups \(c\) over multiple demographic dimensions \(D\) (e.g., age \(\times\) gender \(\times\) education tuples) and to recover coherent marginal turnout tables from these cell-level estimates. There are generally three main data sources available for this: (1) a population table with cell counts \(N_c\); (2) official turnout counts \(V_r\) for regions \(r\); (3) a survey with respondent-level covariates \(X_i\) and outcomes \(O_i\), where \(O_i\in\{0,1\}\) indicates voting intention (or reported turnout) of individual \(i\).

Throughout, we index demographic cells by \(c\in\{1,\dots,C\}\) and regions by \(r\in\{1,\dots,R\}\). Let \(N_c\) denote the population count in cell \(c\), and let \(N_r=\sum_{c\in r}N_c\) denote the eligible population in region \(r\). Official turnout provides counts \(V_r\) (voters) for each region \(r\). In the survey, let \(N_S\) be the number of respondents, and let \(n_c\) be the number of respondents in cell \(c\).

\subsection{MRP baseline (BP)}
Multilevel Regression with Poststratification (MRP) \citep{park_bayesian_2004, ghitza_deep_2013} relies mainly on survey data that is then adjusted to match the population cell counts. Fundamentally, MRP has three parts: (i) a regression model for an individual-level outcome (here turnout) as a function of demographics; (ii) multilevel (hierarchical) priors that partially pool across sparse categories; (iii) poststratification, i.e. aggregating fitted cell probabilities using population counts from the census.

All models in this paper share the same multilevel linear predictor for the latent voting propensity (outcome) of individual \(i\) (with \(i\) belonging to cell \(c\)):
\begin{equation}
\eta^o_i = \beta^o_0 + \sum_{d=1}^D \beta^o_{d,j[i]},
\label{eq:linear_predictor}
\end{equation}
where \(\beta^o_{d,j[i]}\) is the coefficient for the category of covariate dimension \(d\) to which \(i\) belongs. For partial pooling we use simple two-level hierarchical priors: for each categorical dimension \(d\) with categories \(j=1,\dots,J[d]\),
\begin{align*}
\beta^o_{d[j]} &\sim\text{Normal}\big(0, \tau_d^o\big), \quad &\text{for } j = 1,\dots,J[d]\\
\tau_d^o &\sim\text{Half-Normal}(0, \sigma_0^o), \quad &\text{for } d = 1,\dots,D.
\end{align*}
Within each dimension the category effects are constrained to sum to zero (implemented as a zero-sum normal), so that \(\beta^o_0\) is identified as the overall level and each \(\beta^o_{d,j}\) is a deviation from it; the same constraint is used for the selection predictor \eqref{eq:linear_predictor_selection}.

To turn the linear model into a binary prediction, we use a probit\footnote{It is more common to use a logistic link in the literature, but we use probit as we later make use of a bivariate distribution where it is mathematically more convenient. In practice, this does not affect results beyond a scaling factor on the priors.} link, so we model
\begin{equation}
O_i \sim \text{Bernoulli}(\Phi(\eta_i^{o})),
\label{eq:bernoulli_obs}
\end{equation}
where \(\Phi\) denotes the standard normal CDF (used in this role throughout). Poststratification then yields any desired margin by aggregating \(\Phi(\eta_c^{o*})\) with weights proportional to \(N_c\). In particular, for any region \(r\), define the model-implied average turnout as a transformed random variable
\begin{equation}
p_r = \frac{\sum_{c \in r} N_c\,\Phi(\eta^o_c)}{N_r}.
\label{eq:region_poststrat}
\end{equation}
We refer to this model as Basic Probit (BP), as everything described above is shared by all subsequent models.

\subsection{Ghitza-Gelman (GG)}

\citet{ghitza_deep_2013} proposed a post-hoc calibration of MRP estimates to match known aggregates. Their method applies a scalar shift in each region to the estimated cell probabilities so that aggregated totals match official counts. To our knowledge, this is the current state-of-the-art approach to this problem in political science literature, and we thus use it as a benchmark for our models.

\section{Proposed Models}
We propose three new models that build on the existing models in incremental steps.

\subsection{Multilevel ecological inference (EI)}
Our first innovation is a multilevel ecological inference (EI) model that extends the MRP modeling logic to aggregate data. We use the same hierarchical linear predictor to model turnout at the demographic-cell level, and then map these cell-level probabilities to regional turnout via poststratification. Let \(r\) index regions, and write \(c\in r\) if cell \(c\) belongs to region \(r\). Using the poststratified regional mean \(p_r\) from \eqref{eq:region_poststrat}, we write a likelihood for the official turnout totals:
\begin{equation}
V_r \sim \text{Binomial}(N_r, p_r).
\label{eq:margin_obs}
\end{equation}
This yields our EI model. The key conceptual novelty relative to classic EI is that most established EI formulations target an \(R\times C\) table over a single demographic dimension, possibly with additional region-level covariates \citep{king1999, rosen_bayesian_2001, wakefield_ecological_2004, pavia_ecolrxc_2025}. In contrast, we treat turnout as a function of a high-dimensional cross-classification of demographics and use multilevel priors to regularize this otherwise under-identified aggregate-data problem, thereby injecting census structure directly into EI. We believe this formulation holds promise wherever rich census cross-tabs exist---potentially even against state-of-the-art \(R\times C\) models---but such a comparison is outside our scope; here EI serves as a building block and a proxy for what margins data alone can achieve.

\subsection{Joint Poll-and-Margin (PM)}\label{sec:PM}
Our second innovation combines the two preceding models into one. BP and EI are two different observation channels on the \emph{same} population and the \emph{same} latent cell-level turnout field: BP observes individual survey responses, EI observes official regional totals, and both are driven by the outcome predictor \(\eta^o\) of \eqref{eq:linear_predictor}. The Poll-and-Margin (PM) model couples them by letting both likelihoods depend on this single shared predictor---the same intercept, category effects, and hierarchical priors---rather than fitting two separate coefficient vectors.

Because survey turnout reports are typically inflated (e.g., by social desirability), we add a single additive survey-bias term \(\beta^o_{or}\) that shifts \emph{only} the survey predictor, \(\eta_i^{o*} \equiv \eta_i^o + \beta^o_{or}\), leaving the population process \(\eta^o\)---and hence the poststratified margins \eqref{eq:region_poststrat}---unchanged. PM then pairs the survey likelihood \eqref{eq:bernoulli_obs}, evaluated at the shifted predictor \(\eta_i^{o*}\), with the official-margin likelihood \eqref{eq:margin_obs}, both functions of the shared coefficients \(\theta = (\beta^o_0, \{\beta^o_{d,j}\}, \{\tau^o_d\}, \beta^o_{or})\). Treating the two data sources as conditionally independent given \(\theta\), the joint log-posterior is the sum of the BP and EI contributions and the shared prior:
\begin{equation}
\log p(\theta \mid \{O_i\}, \{V_r\}) = \underbrace{\sum_i \log \operatorname{Bern}\!\big(O_i;\, \Phi(\eta^{o*}_i)\big)}_{\text{BP survey term}} + \underbrace{\sum_r \log \operatorname{Bin}\!\big(V_r;\, N_r, p_r\big)}_{\text{EI margin term}} + \log p(\theta) + \text{const}.
\label{eq:pm_logpost}
\end{equation}
PM is thus exactly the additive merger of BP and EI over a common coefficient vector: dropping the survey term recovers EI, and dropping the margin term recovers BP. (Because the survey respondents are a small subset, \(N_S \ll N_r\), of the population already counted in \(V_r\), this conditional independence slightly double-counts them, but the effect is negligible at realistic sampling fractions.)

The two sources contribute complementary, non-redundant information, which is what makes the merger worthwhile. The official margins are near-exact constraints on the poststratified totals \(p_r\), so they pin down the overall \emph{level} of turnout and its variation across regions; the individual-level survey identifies the \emph{shape} of \(\eta^o\) across the demographic cross-classification---the fine within-region contrasts between cells that aggregate data alone cannot resolve, and that EI can regularize only through its priors. The bias term \(\beta^o_{or}\) reconciles the two: from the survey alone it is confounded with the intercept \(\beta^o_0\), but once the margins fix the poststratified level, \(\beta^o_{or}\) absorbs the survey's over-report so that the survey informs cell structure without dragging the overall level away from the official totals.

Unlike the post-hoc calibration of GG, PM fits both sources within one coherent likelihood, so the uncertainty in each propagates into the shared coefficients and hence into every cell-level estimate. In spirit, PM is a turnout-scale instance of the aggregate-plus-individual hierarchical models of \citet{jackson_improving_2006}, with the multilevel census cross-classification providing the shared structure and \(\beta^o_{or}\) absorbing over-reporting.

\subsection{Full Selection (FS)}\label{sec:FS}
Our third proposal is to additionally model selection into the survey. We augment our standard outcome model \eqref{eq:linear_predictor} into a Heckman-style bivariate probit model by modeling it jointly with selection \eqref{eq:linear_predictor_selection} and with correlated errors:
\begin{equation}
\eta^s_i = \beta^s_0 + \sum_{d=1}^D \beta^s_{d,j[i]},
\label{eq:linear_predictor_selection}
\end{equation}
with
\begin{align}
O_i^* &= \eta^{o*}_i + \epsilon_i^o \quad \text{(Outcome)} \label{eq:outcome}\\
S_i^* &= \eta^s_i + \epsilon_i^s \quad \text{(Selection)} \label{eq:selection}
\end{align}
where \((\epsilon_i^o, \epsilon_i^s) \sim \mathcal{N}(0, \boldsymbol{\Sigma})\) and \(\boldsymbol{\Sigma}\) has unit variances and correlation \(\rho\). We observe survey response \(S_i = \mathbb{I}(S_i^* > 0)\) and, if \(S_i=1\), voting intent \(O_i = \mathbb{I}(O_i^* > 0)\).

Two practical issues typically make selection models hard to use in survey work. First, one must be able to model selection even though we do not observe covariates for non-respondents. Under random contact designs (e.g., RDD), we can use the census \(N_c\) to represent the contacted population and infer the demographic tilt of respondents via a multinomial likelihood over cells. We define
\begin{equation*}
\pi_c^s = \frac{\Phi(\eta^s_c)\,N_c}{\sum_{c'} \Phi(\eta^s_{c'})\,N_{c'}}.
\end{equation*}
This allows us to model the likelihood for the number \(n_c\) of survey respondents in cell \(c\) as
\begin{equation}
(n_c)_c \sim \text{Multinomial}(N_{S}, (\pi_c^s)_c).
\label{eq:selection_multinomial}
\end{equation}

Second, classical selection models often rely on an exclusion restriction: a variable that affects selection but not turnout. Designing such instruments into data collection can be costly and requires fine control over the polling process \citep{bailey_countering_2025}, making the correction often infeasible. Our key insight is that combining (i) selected survey outcomes with (ii) selection-free official turnout totals (the election) and (iii) census cell sizes provides a natural source of additional---albeit, as we show below, only partial---identification. Intuitively, the official margins anchor the outcome process even when the survey is tilted, helping to separate selection and outcome effects without a purpose-built instrument.

The calculus needed to properly marginalize the selection component and implement efficient computation is somewhat technical and is presented in \ref{sec:appendix_model_derivation}, but leads to introducing a latent value \(u_i\) per survey respondent, after which the model is
\begin{align*}
u_i \mid \beta^s &\sim \text{Normal}(0,1) \text{ truncated to } (-\eta_i^s, \infty),\\
p^{o\mid s}_i &\equiv \Phi\!\left(\frac{\eta_i^{o*} + \rho u_i}{\sqrt{1-\rho^2}}\right),\\
\qquad
O_i \mid u_i,\beta^o,\rho &\sim \text{Bernoulli}(p^{o\mid s}_i),\\
p_r &\equiv \frac{\sum_{c \in r} N_c\,\Phi(\eta^o_c)}{N_r},\\
V_r &\sim \text{Binomial}(N_r, p_r).
\end{align*}

This additional identification is unfortunately not enough to fully identify both the overreporting bias \(\beta^o_{or}\) and the selection intercept \(\beta^s_0\) as both manifest mainly as increased reported turnout in the survey. This is amplified by the selection effect being nearly linear in most of the domain (which is easy to see via inverse Mills approximation), leading to exactly the classical weak identification issue Heckman models usually suffer from. There are two possible solutions to this problem.

Option one is to drop the overreporting effect and assume all overreporting is caused by selection effects. While unrealistic in the light of existing literature \citep{sciarini-overreporting-heckman, holbrook_social_2010}, it can nevertheless provide a useful approximation in situations where selection bias is expected to dominate, or serve as an upper bound to how strong selection effects could be.

Option two is to set informative priors on the two weakly identified values. This is likely the more promising option, as we can be relatively confident that overreporting bias \(\beta^o_{or}\) is positive and that response rates are below 50\% under most polling regimes. Furthermore, in many situations it should be possible to get a pretty good estimate for response rate with a ratio of respondents to total number of people contacted, which can directly be turned to a very narrow prior for \(\beta^s_0\). This identification can be further helped with a stronger prior on \(\rho\) that simply captures the assumption it is likely to be positive and not too extreme. This is the approach we take in this paper.

\section{Implementation}

We use semi-informative priors for \(\beta\) by setting \(\sigma^o_0, \sigma^s_0 = 1\) and \(\beta^o_0 \sim \text{Normal}(0,1)\). For the other parameters, for reasons listed in \autoref{sec:FS}, we use informative priors, setting over-reporting to \(\beta^o_{or} \sim \text{AsymmetricLaplace}(0,1/3,15)\) so it is 9 times likelier to be positive than not, while keeping the mode at zero; \(\beta^s_0 \sim \text{StudentT}(3,-1.5,0.75)\), which on the response-rate scale puts the center at \(\Phi(-1.5) \approx 6.7\%\) with an 80\% CI of (0.3\%,39.3\%) and significant mass at the tails; \(\rho \sim 2\text{Beta}(6,3)-1\) so the 80\% CI is (-0.1,0.71) with a mode of 0.43. These values are chosen to be informative, prioritizing most expected empirical values while still allowing for full variation.

We implement all models using PyMC \citep{pymc2023} and sample them with the No U-Turn Sampler (NUTS; \citealp{hoffman_no-u-turn_2011}) as implemented in the nutpie sampler. Naive implementations diverge frequently and occasionally fail entirely; reparameterization, log-space computation, and probability clipping resolve most of these issues (details in the companion package, \citealp{salk-turnout-models}).

After these changes, there are still a few divergences, even after increasing the \verb+target_accept+ parameter, and a minority of FS fits additionally show elevated \(\hat{R}\) values indicating incomplete mixing. Both issues are almost entirely confined to the FS model, and their most likely cause is the weak identification discussed in \autoref{sec:FS}: the selection intercept \(\beta^s_0\), the overreporting bias \(\beta^o_{or}\), and the error correlation \(\rho\) jointly form a likelihood ridge along which the sampler can drift, with the informative priors softly regularizing but not eliminating it. Consistent with this explanation, the convergence problems concentrate in settings where selection effects are weak or absent---exactly where \(\rho\) is barely identified and the ridge is flattest---rather than in the challenging high-selection settings the model is designed for (see \ref{sec:appendix_simulation_study} for details). As the following experiments show, the models work well despite these issues, and the comparisons are unchanged when poorly converged runs are excluded; still, FS results with elevated \(\hat{R}\) should be treated with caution.

\section{Extensions}
The modularity of our framework invites several extensions, implemented in the companion package and detailed in \ref{sec:appendix_extensions}: ordinal (Likert-scale) outcomes via an ordered probit; multiple simultaneous polls and margin sets; a beta-binomial margin likelihood with a learned concentration parameter, which softens the near-hard constraints the binomial margin likelihood imposes when the census and turnout frames do not match exactly (evaluated on the Estonian data in \autoref{sec:application}); and a robust Student-\(t\) selection mechanism \citep{MarchenkoGenton, ding2014}.

\section{Simulation Experiments}
\label{sec:simulation}
We conducted a comprehensive simulation study to evaluate model performance under controlled conditions, following the ADEMP framework outlined by \citet{Morris_2019}. The primary aim was to understand when the proposed models perform well and if they have any strong limitations. What follows is a brief summary of methodology and results, with the full details in \ref{sec:appendix_simulation_study}.

\subsection{Data-Generating Process}
We used the Estonian 2021 census as bundled for replication in \cite{salk-turnout-models} (867,608 eligible voters in the cross-tabulation used for simulations) as the known population demographic structure \citep{stateston_rl21303}. We focus on Estonia for pragmatic reasons: it has a relatively small electorate and publicly available, granular multidimensional census tables, giving the simulated populations realistic between-attribute correlations that are complex, non-linear, and otherwise hard to simulate. The dataset contains the counts of demographic cells over five demographic dimensions: age (7 categories), gender (2), education (3), nationality (2), and region (24)\footnote{County-level units (Estonia's 15 \emph{maakonnad}), with Tallinn split into its 8 city districts and Tartu separate---24 in total. We call these ``regions'', matching the model's region index \(r\).}, so every citizen can be placed exactly in these dimensions. To generate a simulated dataset, we generated individual voting and survey response behavior for each individual in the population using a Heckman selection model as described in Equations \eqref{eq:outcome} and \eqref{eq:selection}, with overreporting modeled with additive bias term \(\beta^o_{or}\) as in \autoref{sec:PM}. We then sampled \(N_S\) survey respondents from those individuals for whom \(S_i=1\). Modifications required for more niche scenarios are detailed in \ref{sec:appendix_simulation_study}.

\subsection{Test Scenarios and Performance Measures}
We tested the models (BP, EI, GG, PM, FS) across a range of challenging scenarios drawn from the literature on survey methodology, ecological inference, and selection models (\autoref{tab:test_scenarios}). Each scenario modifies specific aspects of the data-generating process to assess robustness. For each, we generated 11 independent datasets with different random seeds, fitted all models, and compared estimated turnout margins to the true known margins. In total, the eleven scenario families of \autoref{tab:test_scenarios}---each swept over a grid of settings and augmented with a baseline and a no-selection reference---comprise 726 simulated datasets and 3{,}872 fitted models (\(\approx\)320 hours of MCMC). Per-scenario specifications, sampler settings, convergence diagnostics, and results are given in \ref{sec:appendix_simulation_study}.

\begin{table}[ht]
\centering
\caption{Model and Data Test Scenarios}
\label{tab:test_scenarios}
{\footnotesize
\setlength{\tabcolsep}{4pt}
\begin{tabularx}{\linewidth}{>{\RaggedRight\arraybackslash}p{0.58\linewidth} >{\RaggedRight\arraybackslash}X}
\toprule
Scenario & Key References \\
\midrule
Selection Bias / Non-response Bias (varying intercept \(\beta_0^s\)) & \citep{groves_role_2004} \\
Error Correlation (varying \(\rho\)) & \citep{puhani_heckman_2000} \\
Measurement Bias / Over-reporting (adding constant to survey outcome) & \citep{holbrook_social_2010} \\
Aggregation Bias (correlating turnout with group proportion) & \citep{cho_iff_1998, freedman_ecological_1999} \\
Collinearity between selection and outcome processes (correlated \(\beta^s, \beta^o\)) & \citep{puhani_heckman_2000, leung_choice_1996} \\
Non-normal Errors (Student's t and skewed distributions) & \citep{paarsch_monte_1984} \\
Varying Selection and Outcome Effect Sizes (\(\sigma^s_H, \sigma^o_H\)) & -- \\
Interaction Effects (pairwise demographic interactions) & -- \\
Random Noise (asymmetric selection contamination / outcome flips) & -- \\
Varying Sample Size (\(N_S\)) & -- \\
Varying Margin Informativeness (topline, electoral district, region) & -- \\
\bottomrule
\end{tabularx}
}
\end{table}

The main estimand is the full population turnout distribution, i.e. a tensor of the counts of people in each age \(\times\) gender \(\times\) education \(\times\) nationality \(\times\) region \(\times\) voted cell divided by the size of the population. Performance was measured using two distance metrics between estimated and true distributions:
\begin{itemize}
    \item \textbf{Kullback-Leibler (KL) Divergence}: \(D_{KL}(P,Q) = \sum_x P(x)\log(P(x)/Q(x))\), where \(P\) is the true distribution and \(Q\) the estimated distribution over voters and non-voters per cell.
    \item \textbf{Total Variation Distance}: \(D_{TV}(P,Q) = \frac{1}{2}\sum_x |P(x) - Q(x)|\). In our setting of categorical distributions under the discrete metric, this coincides with the Wasserstein-1 (earth mover's) distance.
\end{itemize}
We calculated these for the full margin and for all one- and two-dimensional marginal tables, reporting averages in the latter case.

\subsection{Key Results}

\begin{figure}[H]
    \centering
    \caption{Distance from ground truth under baseline parameters (log scale; linear version in \autoref{fig:a2_default_parameter_metrics}). Here and throughout, points are medians and error bars interquartile ranges across the 11 seeds---seed-to-seed variability, not posterior uncertainty.}
    \label{fig:mb_default_parameter_metrics}
    \includegraphics[width=\textwidth]{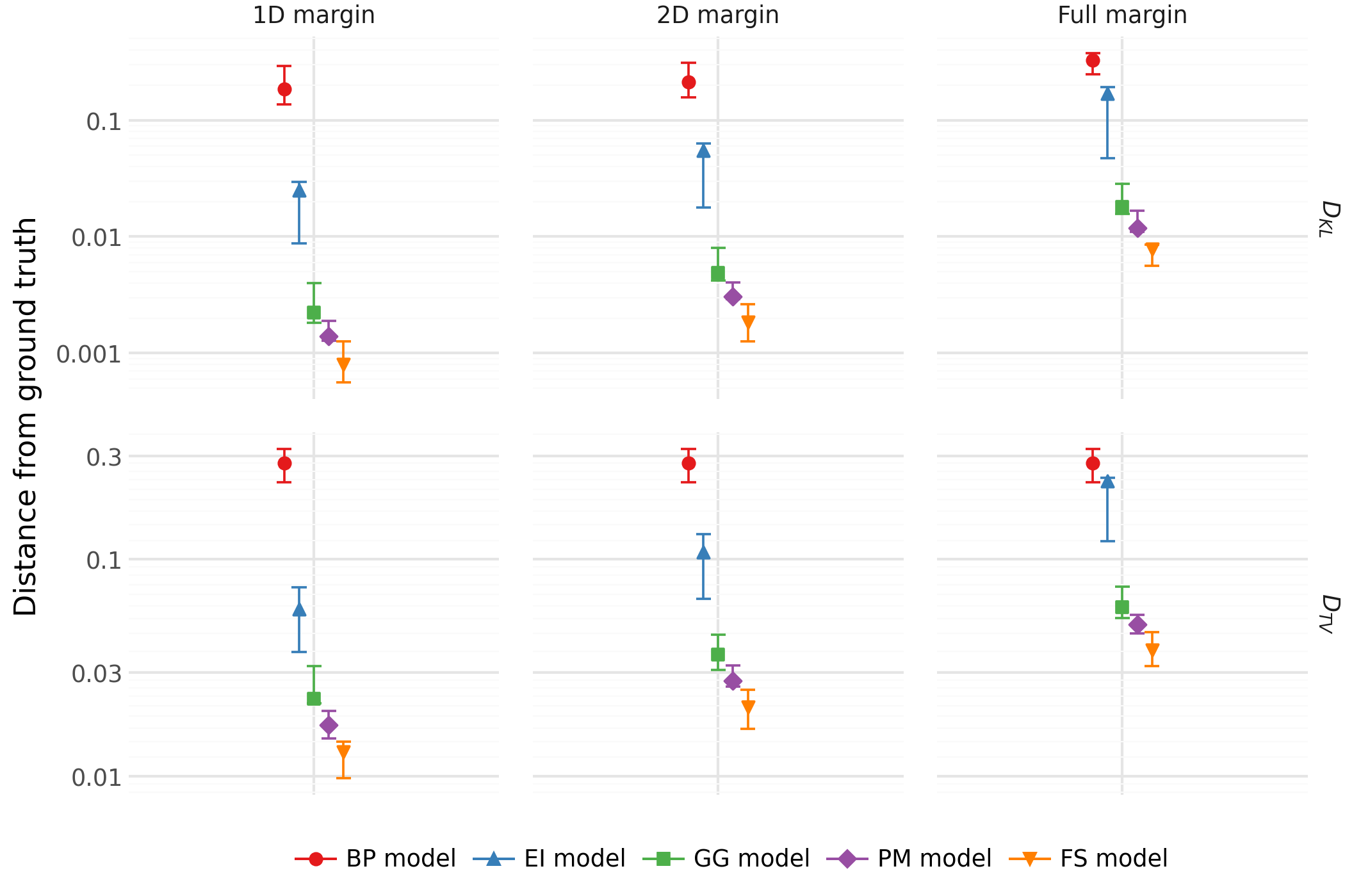}
\end{figure}

The baseline parameters were chosen to reflect moderate selection bias and effect sizes. Coefficients for demographic categories were drawn from normal distributions with zero mean and standard deviations \(\tau_k^s\) and \(\tau_k^o\), which themselves followed half-normal distributions with \(\sigma^s_H=\sigma^o_H=0.5\). The error correlation \(\rho\) was set to 0.5, reflecting empirical evidence that those more likely to respond are also more likely to vote \citep{groves_role_2004}. A survey sample of size \(N_S = 1000\) was drawn uniformly from the subpopulation with \(S_i=1\). Selection intercept was fixed to \(\beta_0^s=-1\) to simulate significant selection bias (with a selection rate of around 21\%) and outcome intercept was fixed to \(\beta_0^o=0\) to simulate turnout of around 50\%. Survey reports additionally shift the latent threshold by \(\beta^o_{or}=0.3\) relative to the true population model (overreporting). Unless a scenario varies margin informativeness, all margin-using models (EI, GG, PM, FS) are given region-level turnout margins (all 24 regions). Results for the baseline case are shown in \autoref{fig:mb_default_parameter_metrics}.

Under baseline parameters, FS performs best, followed by PM, GG, EI, and BP. The median KL divergence from the ground truth of the FS model was 57\% lower than that of the GG model (bootstrap 95\% CI over seeds \([43\%,79\%]\)), and the median total variation distance was 36\% lower in the same comparison. Because all models are fit to the same 11 simulated datasets, these comparisons can be made paired (per seed), which is both more powerful and robust to seed-to-seed difficulty. On a per-seed basis FS achieves lower KL than GG on 10 of 11 seeds (Wilcoxon signed-rank \(p=0.003\)) and PM beats GG on all 11 seeds (\(p=0.001\)); FS's edge over PM is smaller and only marginally significant (8 of 11 seeds, \(p=0.054\)), consistent with our recommendation to report both. The basic models (BP, EI) had significantly wider confidence intervals, indicating greater uncertainty.

Beyond point accuracy, we also assessed whether the models' posterior uncertainty is calibrated---a central advantage we claim for the joint Bayesian models over the two-step GG procedure. On the baseline scenario, PM and FS attain near-nominal 90\% coverage of the true margin values while producing posterior credible intervals three to five times narrower than GG's, so their lower point error is not bought with overconfident intervals; the survey-only BP model, by contrast, is severely overconfident (near-zero coverage). Detailed coverage results are reported in Appendix~\ref{sec:a2_coverage}.

We note that under most scenarios the data are generated from the same bivariate-probit structure that FS assumes, so these results represent a best case for FS; the noise and non-normal-error scenarios probe departures from this assumption.

The model ordering of the baseline case remained stable for most of the other test scenarios with only minor deviations in rankings. Most notably, BP outperforms EI when selection bias is absent or very small, consistent with BP leveraging richer (and in that case nearly unbiased) individual-level data. Average results for all scenarios are provided in \ref{sec:appendix_simulation_study}.

\begin{figure}[H]
    \centering
    \caption{Kullback-Leibler Divergence With Different Degrees of Censoring}
    \label{fig:mb_selection_bias_intercept_metrics}
    \includegraphics[width=\textwidth]{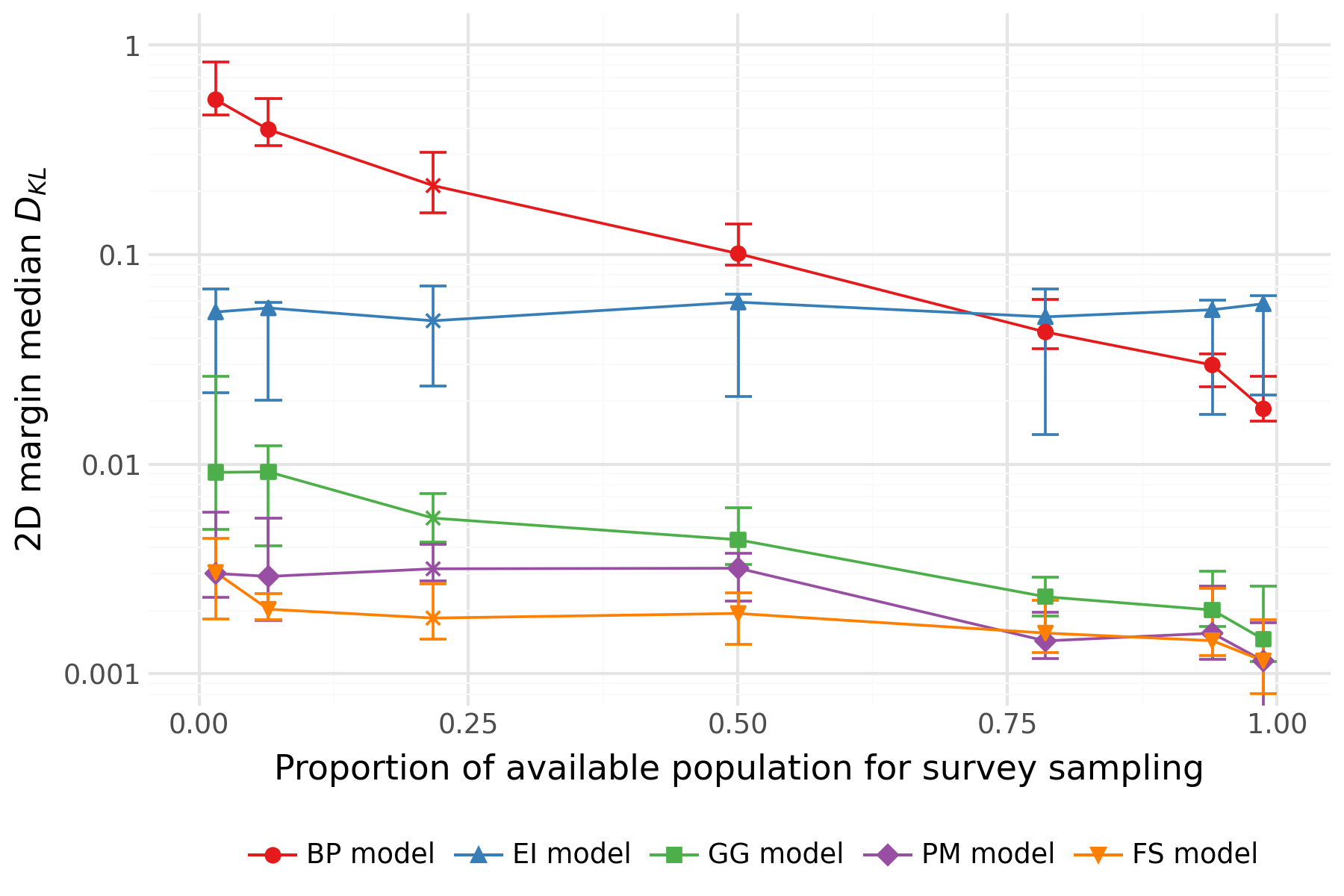}
\end{figure}

As can be expected, the joint models (PM and FS) showed an especially strong advantage over GG under high selection bias (see \autoref{fig:mb_selection_bias_intercept_metrics}). When less than 1\% of the population was accessible for polling (simulated via a selection intercept \(\beta_0^s = -3\)), both PM and FS still provided about as good estimates on the 2D margins as they did in the baseline case with \(\sim\)21\% accessibility. While FS maintains an edge over PM on the full joint distribution (see \ref{sec:appendix_simulation_study}), they perform effectively identically on the 2D margins at this extreme level of non-response. This convergence is partly an artifact of the informative prior on \(\beta_0^s\): when the true intercept lies deep in the prior's tail (as at \(\beta_0^s=-3\), against a prior centered at \(-1.5\)) the posterior is pulled upward, so FS underestimates the severity of selection and applies a correspondingly muted correction, behaving much like PM, and emphasizing the dependence on informative priors. As the accessibility increases to the baseline level of \(\sim\)21\%, the intercept is recovered accurately and FS separates itself, clearly outperforming PM across all metrics and margin levels. As the proportion of the population willing to answer surveys increases further towards 100\%, all models improve, and differences between them diminish. This demonstrates the joint models' ability to correct for severe non-response, with FS providing the most robust performance across moderate to severe selection bias.

As the correlation \(\rho\) between selection and outcome errors increased from 0 to 1, all models relying on survey data found estimation more difficult. BP, with no access to aggregate margins, degraded fastest of all; among the margin-using models, GG's performance degraded most rapidly, while the FS model maintained the lowest distance metrics. With zero correlation (no selection bias), all models except EI performed similarly well, as expected.

With smaller survey samples (\(N_S < 500\)), the more complex models (PM, FS) showed larger variance but maintained better accuracy than the simpler models; as \(N_S\) increased, FS showed the steepest proportional reduction in median KLD (the figure uses a log scale). Similarly, providing more granular aggregate margins (e.g., region-level instead of electoral-district-level) improved estimates for all models that use them, with EI and FS showing the largest proportional gains.

None of the models had any issues with the datasets generated to simulate potentially problematic cases of coefficient collinearity or aggregation bias. All models are also reasonably robust to errors being Student T or even skewed Student T.

However, the FS model exhibits a critical sensitivity not captured by these controlled scenarios. When the simulated data include random selection noise—where the selection process itself has stochastic components beyond the modeled demographic predictors—the edge FS has over PM and GG disappears already at light selection noise (5\%). Two effects contribute to this. First, FS's advantage rests on the modeled selection mechanism being accurate, and the noise breaks that assumption. Second, the injected noise makes the realized sample less selective overall, so there is simply less selection bias left to correct: the models without a selection correction (GG, PM) in fact \emph{improve} as noise increases, while FS merely stays level. Either way, the practical conclusion stands: the advantage of FS should only be expected when the survey's selection mechanism is reasonably well captured by the model.

\section{Application: Estonian 2023 Parliamentary Elections}
\label{sec:application}
We applied the models to estimate turnout demographics for the 2023 Estonian parliamentary election, held March 5, 2023. Official turnout was 63.5\% including voters abroad and 70.7\% after adjusting to match the domestic census frame \citep{estonia_rk2023_results}. Data included: (1) the replication census cross-tabulation (867,608 eligible voters) cross-classified by age (7 categories), gender (2), education (3), nationality (2), and region (24) \citep{stateston_rl21303}; (2) official region-level turnout counts \(V_r\) from the State Electoral Office \citep{estonia_rk2023_results, estonia_elections_opendata}; (3) a two-wave CATI survey conducted by Norstat in February and March 2023 (N=1,192 after filtering) with voting intention.

\subsection{Data Processing Challenges}
Official turnout counts include voters living abroad (\(\sim\)5-10\% of eligible voters, turnout \(\sim\)10.7\%), who are absent from the census and survey \citep{estonia_rk2023_results, estonia_elections_opendata}; as their region of registration is unknown, we excluded them from the counts \(V_r\) and redistributed their votes proportionally across electoral districts. Excess votes from people voting outside their own electoral district (0.5-3\%) were likewise distributed proportionally among constituent regions. These adjustments align the census population with the turnout denominators.

The survey was originally conducted as a quota sampling phone survey followed by an extra sample from a web panel to make it more representative. We drop the web respondents from the analysis as their selection model is far from random contact. Furthermore, to make the phone sample match the random contact assumption, we limit the selection model \eqref{eq:selection_multinomial} to the first 80\% of respondents of each wave, as major quota get filled up only past that point. The survey suffered from significant over-reporting: 91.6\% of respondents stated they intended to vote or had already voted, compared to the 70.7\% adjusted turnout rate used for modeling. This matches known patterns of social desirability bias \citep{holbrook_social_2010}.

\subsection{Results}
We fitted all models (BP, EI, GG, PM, FS) using region-level turnout margins and evaluated estimates against known age and gender turnout margins (other margins were not available for validation).
\begin{figure}[H]
    \centering
    \caption{Estonia Turnout Model Distance Metrics}
    \label{fig:estonia_turnout_metrics}
    \includegraphics[width=0.9\textwidth]{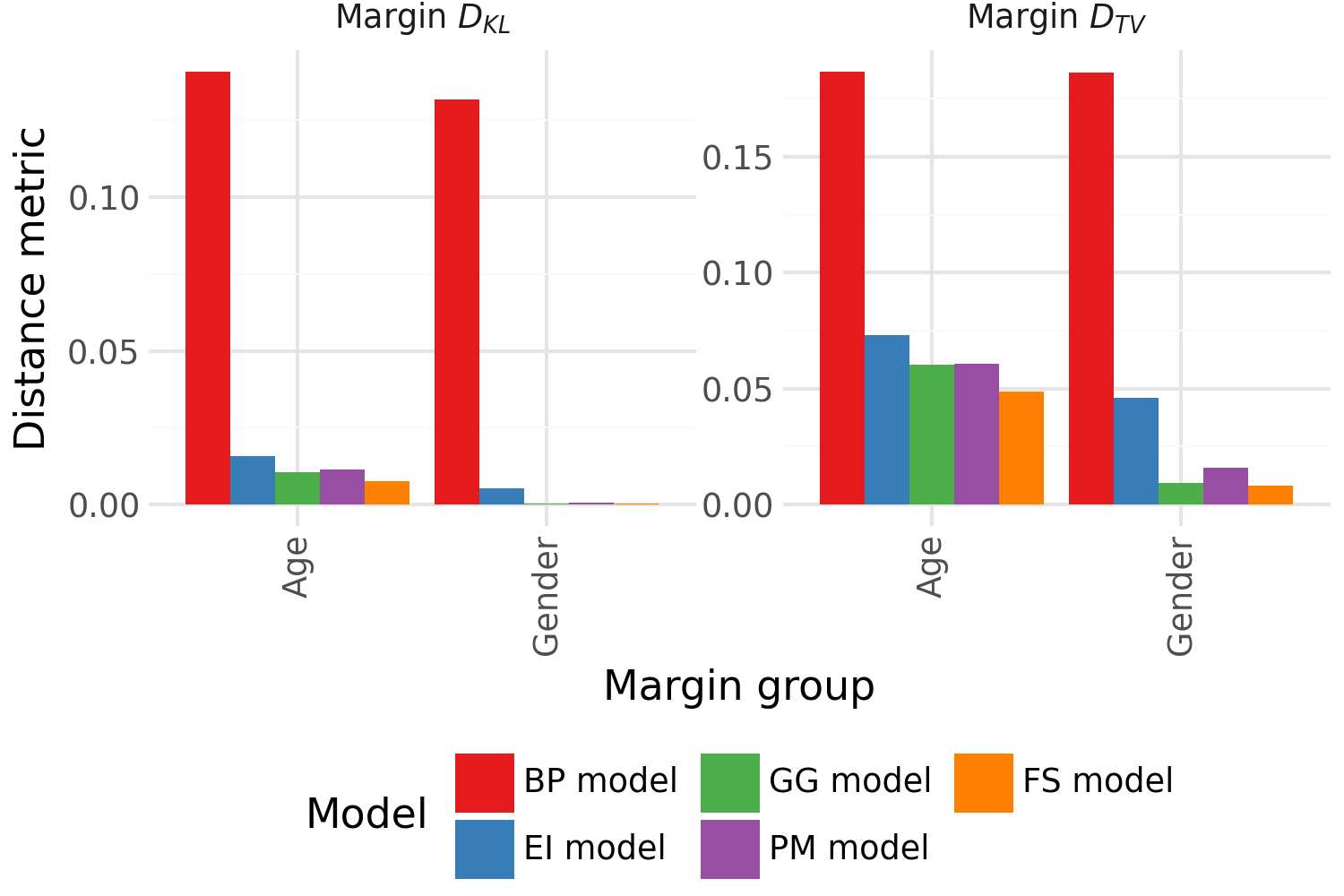}
\end{figure}

Results in \autoref{fig:estonia_turnout_metrics} show the BP model estimates were furthest from the known margins, as expected given its reliance on biased survey data. The other models perform very similarly however, at least on averages, with FS model having a very slight edge that is likely not statistically significant.

\begin{figure}[H]
    \centering
    \caption{Estonia Turnout Model Posterior Margins}
    \label{fig:estonia_1d_margins}
    \includegraphics[width=\textwidth]{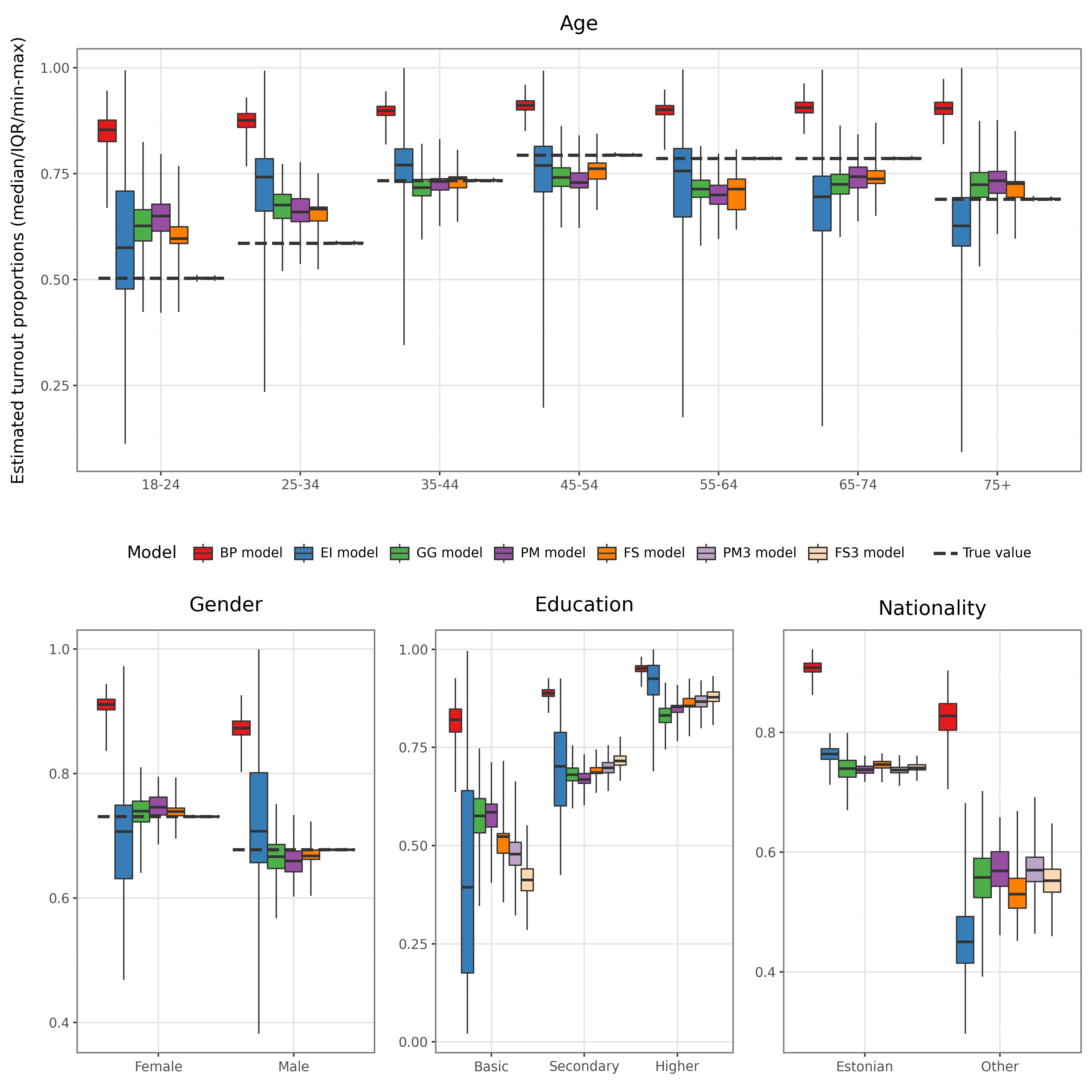}
\end{figure}

To look at what the models imply for the margins more precisely, we fit two additional models: PM and FS with all three known topline margins given as input instead of just the region-level one (labeled PM3 and FS3, respectively). Results are displayed on \autoref{fig:estonia_1d_margins}.

\begin{figure}[H]
    \centering
    \caption{Estonia Turnout Model Education by Nationality Margin}
    \label{fig:estonia_2d_margin}
    \includegraphics[width=\textwidth]{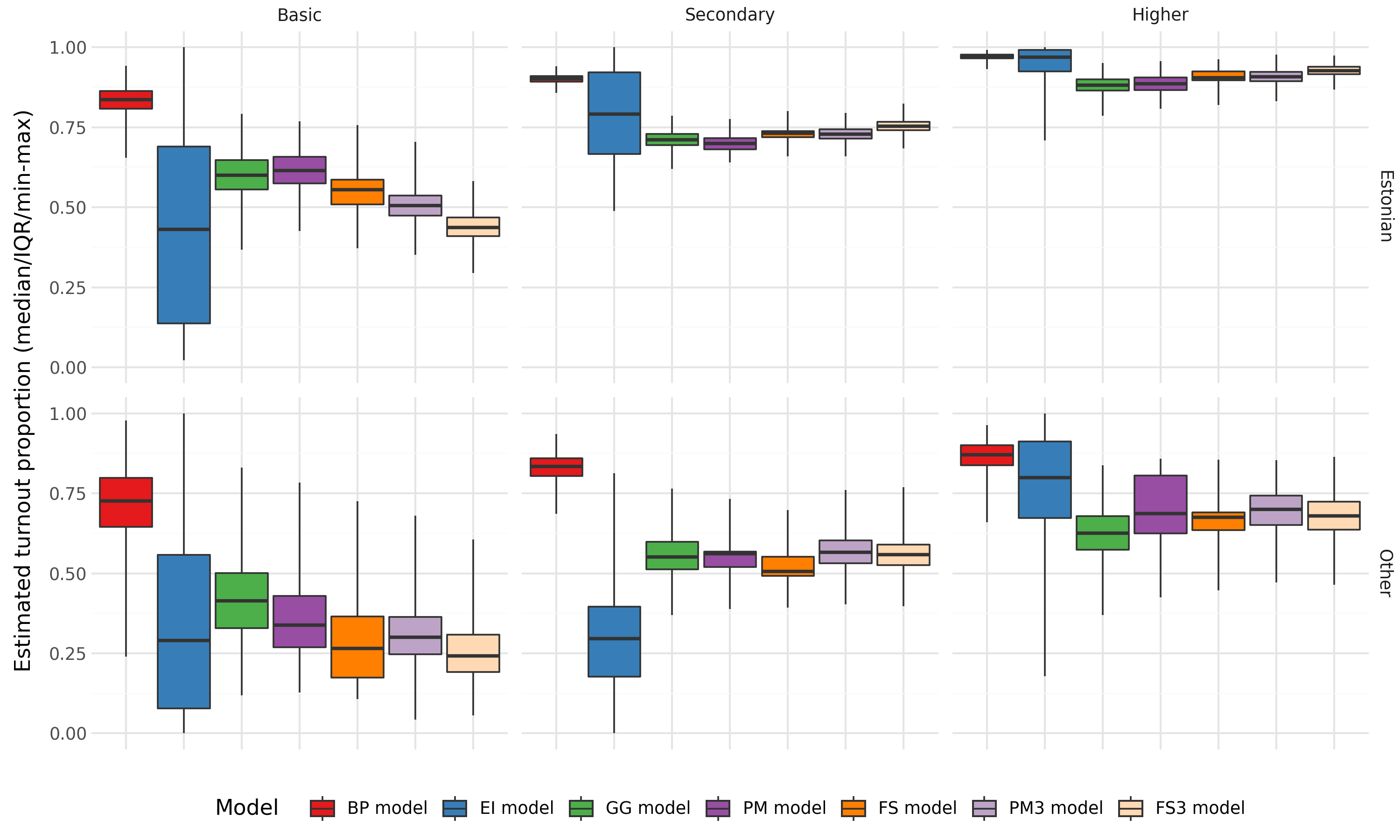}
\end{figure}

In the posterior margins, EI stands apart, unsurprisingly: working from margins alone, its posteriors are extremely wide, often spanning nearly the full 0--100\% range. The more advanced models (GG, PM, FS) all converge to very similar values, with FS tilting away from the others slightly. This tilt is further strengthened in PM3 and FS3, indicating adding more data supports the tilt direction FS finds. Looking at the distribution of $\rho$ in the posteriors, it seems a large part of this effect is due to FS3 gaining more confidence in the selection effect, with a modal $\rho$ of ~ 0.6 compared to ~0.4 for the FS model. The same patterns persist looking at the estimates for 2D margins such as nationality by education (\autoref{fig:estonia_2d_margin}).

The posterior margins also make the frame-mismatch issue concrete: with the binomial margin likelihood, the age-margin posteriors of the joint models systematically miss the known values (e.g.\ overestimating turnout in the youngest group), because the near-hard region constraints force any census/turnout frame mismatch into the outcome coefficients. Refitting PM and FS with the beta-binomial margin likelihood of \ref{sec:appendix_extensions} (learned concentration $\approx$430--490) modestly improves the held-out age margin (mean absolute error 0.066$\to$0.063 for PM and 0.053$\to$0.051 for FS, with rising interval coverage) but does not close the gap---so the miscoverage is only partly a frame-mismatch artifact and partly reflects genuine model limitations.

These results all come with an important caveat however. FS model only performs as well as it does once informative priors are used to identify it as described in \autoref{sec:FS}. Indeed, running FS with a flat prior on \(\rho\) and weakly informative priors on \(\beta^o_{or},\beta^s_0\) leads to a modal \(\rho\) value of zero and a posterior distribution of turnout nearly indistinguishable from PM. Considering the informative priors are grounded in previous research and quite wide, we believe their use to be fully justified, but it is crucial to understand that they do play an important role in the actual functioning of the model in the empirical setting. It is worth stressing that PM model does not have this issue, not relying on priors for identification and being much less sensitive to them.

We would also like to emphasize that currently, the method has only been tested on exactly one election in one country, so while the results are promising, they should not be over-interpreted to mean FS will always outperform the other models. Even in the simulation data, there are cases where PM gets closer to the truth simply by chance, and this is likely to happen much more commonly with real data, as PM is simply more robust and relies on fewer assumptions. As such, it is our recommendation that if this methodology is used, results from both PM and FS model be reported and contrasted with each other before any further analysis.

\section{Discussion}
We propose three new models for differential turnout estimation and evaluate them against established baselines, including the state-of-the-art Ghitza--Gelman (GG) calibration approach. All three proposed models rely on two key inputs: granular multi-way census cell counts and official turnout statistics. This makes them primarily feasible for post-election analysis in settings where such data are available.

Our multilevel ecological inference (EI) model applies MRP-style hierarchical structure to the classical EI problem; while we believe the formulation is attractive when granular census data exist, a systematic comparison to existing EI approaches remains outside our scope.

Our joint Poll-and-Margin (PM) model provides a principled single-stage alternative to the two-step GG procedure by fitting survey and aggregate margins within one coherent Bayesian model. Across the simulation study PM consistently outperformed GG---its median KL divergence from the ground truth was roughly 30\% lower, and it was the more accurate of the two in 94\% of paired (per-seed) comparisons and in every scenario's median---whereas on the limited age and gender margins available for validation in the Estonian application the two performed similarly. Combined with its cleaner probabilistic foundations and uncertainty propagation, this makes PM a compelling replacement for GG and our recommended default for practical applications.

Finally, the Full Selection (FS) model extends PM with a Heckman-style selection mechanism to address non-ignorable survey non-response. Use of this model assumes the data come from a random-contact survey and, ideally, that some information is available on the response rate. When these assumptions are met and the identifying priors are well-grounded, the model can provide modest improvements in per-group turnout estimates; however, empirical gains can be small and are sensitive to prior choice. We therefore recommend FS always be used in conjunction with PM and results contrasted with it; used responsibly, it can also help separate over-reporting from selection effects.

Future work should also explore richer modeling choices for \(\eta^o\) and \(\eta^s\): structured priors \citep{Si_2020, Gao_2021} could stabilize estimation in sparse settings, while machine-learning approaches such as BART \citep{chipman_bart_2010} might capture complex non-linearities in data-rich ones.

Perhaps the most significant potential extension, however, lies in generalizing the framework to multinomial outcomes. This would allow the model to estimate not just who votes, but who they vote for, effectively unifying turnout and vote choice estimation into a single coherent framework for multi-party elections. While this generalization is straightforward for the PM and EI models, adapting the FS model's selection correction is considerably more involved and will need to be left for a follow-up paper.

Finally, as our empirical application covers a single election, the models---FS especially---should be tested on a wider range of datasets to better understand their sensitivity to real-world selection noise and to refine the informative priors required for identification.

\paragraph{Acknowledgments}

The authors made use of OpenAI ChatGPT 5.2 to assist with drafting and editing this article during 19 Jan - 23 Feb 2026 to help reorder and rephrase text for better readability and notational consistency. Different LLMs were used to assist in building the software for the simulations during Jul 2025--Jan 2026, most notably Anthropic Claude Sonnet 3.5 and OpenAI ChatGPT 5.2, but all final code was manually reviewed by the authors. The authors are entirely responsible for the scientific content of the paper and for ensuring that the paper adheres to the journal’s authorship policy.

We thank Jaan Erik Pihel for comments on the draft paper and Andres Võrk and Kaur Lumiste for constructive feedback on the simulation study.

\paragraph{Funding Statement} This research was independently funded by SALK.

\paragraph{Competing Interests} The authors declare no competing interests.

\paragraph{Data Availability Statement}
Both the data and the code required for replication are available at \cite{salk-turnout-models}. All five models (BP, GG, EI, PM, FS) are published under the MIT licence on the Python Package Index as \verb+salk_turnout_models+.

\paragraph{Author Contributions}
Conceptualization: T.J.; M.N. Methodology: M.N. Formal analysis: M.N.; R.R. Software: R.R.; M.N. Data curation: R.R.; M.N. Data visualisation: R.R. Writing (original draft): M.N.; R.R. Funding acquisition: T.J. All authors approved the final submitted draft.


\printbibliography
\appendix

\renewcommand\thesubsection{\arabic{section}.\arabic{subsection}}
\renewcommand\thesubsubsection{\thesubsection.\arabic{subsubsection}}

\section{Derivation of Full Selection model} \label{sec:appendix_model_derivation}

In this section we will perform a step-by-step derivation of the FS model.

In the main paper body, we specify models using sampling statements (e.g., $V_r \sim \text{Binomial}(\cdot)$, $O_i \sim \text{Bernoulli}(\cdot)$, $u_i \sim \text{TruncatedNormal}(\cdot)$). In this appendix, we switch to ``log-likelihood'' notation for compactness in the algebra. These are equivalent: given a sampling model, the (joint) likelihood is simply the joint probability mass/density of the observed data as a function of the parameters. Under conditional independence across individuals (given parameters), this joint probability factorizes as a product over $i$, and the corresponding log-likelihood is the logarithm of that product. Concretely, for the bivariate probit selection model used here,
\[
P(S_i=0)=\Phi(-\eta_i^s),\quad
P(S_i=1,O_i=1)=\Phi_2(\eta_i^o,\eta_i^s;\rho),\quad
P(S_i=1,O_i=0)=\Phi_2(-\eta_i^o,\eta_i^s;-\rho),
\]
and multiplying these per-individual probabilities over $i$ yields the likelihood expressions below.

The bivariate normal cumulative distribution function $\Phi_2$ is defined as:
\[
\Phi_2(x_1, x_2; \rho) = \int_{-\infty}^{x_1} \int_{-\infty}^{x_2} \phi_2(u, v; \rho)  dv  du
\]
where $\phi_2$ is the bivariate normal probability density function:
\[
\phi_2(u, v; \rho) = \frac{1}{2\pi\sqrt{1-\rho^2}} \exp\left\{ -\frac{1}{2(1-\rho^2)} \left( u^2 - 2\rho uv + v^2 \right) \right\}.
\]

A key identity for computational purposes expresses the bivariate CDF in terms of univariate integrals:
\begin{equation} \label{eq:bivariate-identity}
\Phi_2(x_1, x_2; \rho) = \int^{\infty}_{-x_2} \Phi\left( \frac{x_1 + \rho u}{\sqrt{1-\rho^2}}\right) \phi(u) du
\end{equation}
where $\phi(\cdot)$ is the standard normal PDF and $\Phi(\cdot)$ is the standard normal CDF.

The complete-data likelihood (i.e., the joint probability of the observed indicators $(S_i,O_i)$ under the bivariate probit sampling model) can be written as:
\begin{align*}
\mathcal{L}(\beta^o, \beta^s, \rho) = & \prod_{\{i: S_i = 0\}} \Phi \left( -\eta_i^s \right) \\
& \times \prod_{\{i: S_i = 1, O_i = 1\}} \Phi_2 \left( \eta_i^o, \eta_i^s, \rho \right) \\
& \times \prod_{\{i: S_i = 1, O_i = 0\}} \Phi_2 \left( -\eta_i^o, \eta_i^s, -\rho \right),
\end{align*}
where $S_i$ indicates selection (response to poll) and $O_i$ indicates turnout intention.

We then use data augmentation to introduce latent variables $u_i \sim \mathcal{N}(0,1)$ for each respondent to demarginalize the bivariate normal distribution. Using identity \eqref{eq:bivariate-identity}, we obtain the augmented likelihood:
\begin{align*}
\mathcal{L}_{\text{aug}}(\beta^o, \beta^s, \rho, \{u_i\}) = & \prod_{\{i: S_i = 0\}} \Phi \left( -\eta_i^s \right) \\
& \times \prod_{\{i: S_i = 1, O_i = 1\}} \Phi \left( \frac{\eta_i^o + \rho u_i}{\sqrt{1-\rho^2}}\right) \\
& \times \prod_{\{i: S_i = 1, O_i = 0\}} \Phi \left( \frac{-(\eta_i^o + \rho u_i)}{\sqrt{1-\rho^2}}\right) \\
& \times \prod_{\{i: S_i = 1\}} \mathbb{I}[u_i > -\eta_i^s] \phi(u_i).
\end{align*}

such that marginalizing over $\{u_i\}$ recovers the original likelihood:
\[
\int \mathcal{L}_{\text{aug}}(\beta^o, \beta^s, \rho, \{u_i\}) d\{u_i\} = \mathcal{L}(\beta^o, \beta^s, \rho).
\]

In our case, we are interested in the joint model only for the polling data for which $S_i=1$. We therefore need to condition on that being the case for all $i$, which just means dividing by $P(\forall i : S_i = 1 ) = \prod_i \Phi(\eta_i^s)$, yielding 
\begin{align*}
\mathcal{L}_{\text{aug}}(\beta^o, \beta^s, \rho, u \mid \{S_i=1\}) = & \prod_{\{i\}} \Phi \left( \frac{(2O_i-1)(\eta_i^o + \rho u_i)}{\sqrt{1-\rho^2}}\right) \\
& \times \prod_{\{i\}} \mathbb{I}[u_i > -\eta_i^s]\frac{\phi(u_i)}{\Phi(\eta_i^s)}.
\end{align*}

Considering $\mathbb{I}[u_i > -\eta_i^s]\frac{\phi(u_i)}{\Phi(\eta_i^s)}$ is the density function for Truncated Normal, this is equivalent to just changing the prior for $u_i$ to a much cleaner
\[
u_i \sim \text{Normal}(0,1) \text{ truncated to } (-\eta_i^s, \infty),
\]
while the remaining likelihood expressions above correspond to the following sampling statements
\[
p_i \equiv \Phi\!\left(\frac{\eta_i^o + \rho u_i}{\sqrt{1-\rho^2}}\right),
\qquad
O_i \mid u_i,\beta^o,\rho \sim \text{Bernoulli}(p_i).
\]

\graphicspath{{figures/}}

\section{Simulation study details} \label{sec:appendix_simulation_study}

This appendix provides compact details of the synthetic-data simulation study summarized in the main text (ADEMP framework; \citealp{Morris_2019}).

\subsection{Aim}
We evaluate when the proposed joint models (PM and FS) improve demographic turnout estimation relative to standard baselines, under controlled violations common in survey and EI settings (selection bias, misreporting, aggregation bias, and mis-specification).

\subsection{Synthetic data generation process}
We treat the Estonian census cross-tabulation (age, gender, education, nationality, region) as the fixed population structure, preserving realistic correlations between covariates. For each individual \(i\), we generate latent selection and turnout propensities via a Heckman-style bivariate probit:
\begin{align*}
S_i^* &= \eta_i^s + \varepsilon_i^s, \qquad S_i^P = \mathbbm{1}(S_i^* \ge 0)\\
O_i^* &= \eta_i^o + \varepsilon_i^o, \qquad O_i^P = \mathbbm{1}(O_i^* \ge 0),
\end{align*}
with linear predictors \(\eta_i^s = \beta_0^s + \sum_k \beta^s_{k[i]}\) and \(\eta_i^o = \beta_0^o + \sum_k \beta^o_{k[i]}\) using the same categorical covariates as the fitted models. Errors follow
\[
(\varepsilon_i^s,\varepsilon_i^o) \sim \mathcal{N}\!\left(\mathbf{0},\begin{pmatrix}1&\rho_{\varepsilon}\\ \rho_{\varepsilon}&1\end{pmatrix}\right).
\]
Category effects use partial pooling with sum-to-zero constraints within each covariate, matching the multilevel structure described in the main text. After de-meaning each draw of category effects within a covariate, we rescale them by $\sqrt{J/(J-1)}$ for $J$ categories so that the simulation respects the intended variance of the (approximate) sum-to-zero random-effects construction.

A survey of size \(N_S\) is then drawn uniformly from the selected subpopulation \(\{i : S_i^P=1\}\), yielding respondent-level covariates and reported turnout \(O_i^S\). In scenarios with measurement error, \(O_i^S\) differs from \(O_i^P\) by shifting the reporting threshold (over-reporting).

\subsection{Experimental design}
For each scenario, we generate 11 independent datasets (different random seeds), fit all models (BP, EI, GG, PM, FS), and compare estimated turnout margins to the known population margins. The scenario set is listed in Table~\ref{tab:test_scenarios} in the main text; we vary one mechanism at a time around the baseline parameters reported in the main text (Simulation Experiments section).

The same 11 base seeds are reused across all scenarios. From each base seed we deterministically derive separate sub-seeds for the three stochastic stages of data generation---the draw of demographic coefficients, the individual-level Heckman errors, and the survey respondent sample---so that any stage a scenario leaves unchanged reproduces exactly the same draws as the baseline. Comparisons across scenarios, and between a scenario and the baseline, are therefore matched at the seed level, isolating the effect of the manipulated mechanism from seed-to-seed sampling variability. When a scenario alters the generative process itself (e.g., a heavier-tailed error distribution, added interaction terms, or a different coefficient scale) the affected draws necessarily diverge; the shared seeds keep everything else as closely matched as the modified code path allows.

\subsection{Model fitting}
Priors follow the Implementation section of the main text: category scales use $\mathrm{HalfNormal}(0,1)$ for both processes; $\beta_0^o\sim\mathcal{N}(0,1)$; $\beta_{\mathrm{or}}^o$ is given an asymmetric Laplace prior favoring positive overreporting; $\beta_0^s$ a Student-$t$ prior centered around low overall response probability; and $\rho$ a transformed Beta prior on $(-1,1)$. Sampling uses MCMC with 2 chains, 800 tuning steps, and 500 draws per chain; model likelihoods are as specified in \ref{sec:appendix_model_derivation}. By default, models were sampled with a NUTS \texttt{target\_accept} parameter of 0.9. Any run with more than 25 divergences was re-run with \texttt{target\_accept} set to 0.99, repeating the re-run if excessive divergences persisted. The results reported below use the run with the fewest divergences for each model--dataset pair; all retained runs have 25 or fewer divergences.

We also monitored Gelman--Rubin $\hat{R}$ statistics. For the full grid we retained the mean $\hat{R}$ across parameters for each fit; we report against a lenient threshold of 1.1, but note this is an average and can mask a single poorly mixing parameter. A minority of runs showed unacceptably high values: 205 of the 3872 fitted models (5.3\%) had a mean $\hat{R}$ above 1.1 (129 above 1.2), and re-running with higher \texttt{target\_accept} did not consistently improve these convergence warnings, even when divergences were eliminated. The pattern of these failures is informative: 91\% of them are FS fits, and they concentrate in conditions with weak or absent selection effects---all 11 FS fits in the no-selection scenario, and 37 of 44 FS fits in the censoring scenario with $\beta_0^s \ge 0$---consistent with the weak-identification ridge between $\beta_0^s$, $\beta^o_{or}$, and $\rho$ discussed in the main text, which is flattest when the data contain little selection signal for $\rho$ to identify.

To characterize the mixing problems more precisely than a parameter-averaged $\hat{R}$ allows, we refit the baseline scenario (all five models, 11 seeds) retaining the full posteriors and computed the stricter rank-normalized split-$\hat{R}$ of \citet{vehtari_rank-normalization_2021}. The diagnostics confirm that FS mixes materially worse than the other models but that the problem is diffuse rather than catastrophic: across the 11 baseline seeds FS has a median worst-parameter split-$\hat{R}$ of 1.12 (up to 2.24 on the worst seed) with about 47\% of parameters exceeding the strict 1.01 threshold and a minimum bulk-ESS as low as $\sim$3, whereas BP and GG stay near 1.02 with $<2\%$ of parameters above 1.01 and bulk-ESS in the hundreds (PM and EI sit in between, EI also showing elevated values from its aggregate-only identification). The practical takeaway is that FS has genuine, widespread mild non-convergence concentrated on the weakly identified selection ridge and its many mildly correlated coefficients, yet still recovers the turnout distribution well: the poststratified estimands depend on $\Phi(\eta^o_c)$, which is far better identified than the ridge itself. Consistent with this, the reported comparisons are robust to the convergence warnings---excluding baseline FS runs with mean $\hat{R} > 1.1$ (2 of 11 seeds) leaves the median full-table $D_{KL}$ at 0.0082 versus 0.0078 with all seeds, and does not change the model ordering---and the interval-calibration results below (\autoref{sec:a2_coverage}) show that the FS posterior intervals remain usefully calibrated (near-nominal 90\% coverage) despite the elevated $\hat{R}$.

\subsection{Running the simulation}
The simulation was run on a personal laptop of M.\ Niitsoo with an AMD Ryzen 7 processor using PyMC (v5.28.4), PyTensor (v2.38.2), and the Nutpie sampler (v0.16.6). The full scenario grid (all rows of Table~\ref{tab:test_scenarios}, $11$ seeds per setting, and five focal models plus a robust-$t$ FS variant) is large, and wall-clock time is dominated by MCMC. The total sequential MCMC sampling time across all models in the grid (including the 0.99 \texttt{target\_accept} re-runs) was approximately 320 hours.

Per-fit times give a sense of the computational cost of the added modeling structure. In the baseline scenario (\(N_S=1000\), region margins, no interactions), the median sampling time per fit was 37\,s for BP, 62\,s for GG, 113\,s for PM, 132\,s for FS, and 174\,s for EI; across the whole grid (which includes larger samples, richer margins, and interaction variants) the corresponding medians were 70\,s, 93\,s, 163\,s, 255\,s, and 213\,s. Thus the selection correction in FS roughly doubles the cost of the baseline probit model, which is modest relative to its accuracy gains, while EI is comparatively expensive because it must identify the outcome process from aggregate data alone.

\subsection{Performance measures}
Let \(P\) be the true distribution over the full turnout table (voters/non-voters by demographic cell), and \(Q\) an estimated distribution from a fitted model (after poststratification).
\begin{itemize}
    \item \textbf{Kullback--Leibler divergence} \citealp{kullback_information_1951}:
    \[
    D_{KL}(P,Q)=\sum_{x\in\chi} P(x)\log\!\left(\frac{P(x)}{Q(x)}\right).
    \]
    \item \textbf{Total variation distance} (which for categorical distributions under the discrete metric coincides with the Wasserstein-1 / earth mover distance):
    \[
    D_{TV}(P,Q)=\frac{1}{2}\sum_{x\in\chi}\lvert P(x)-Q(x)\rvert.
    \]
\end{itemize}
We report these for the full table and for all one- and two-way margins (averaged across margins). Uncertainty bands in the plots reflect variability across the 11 generated datasets (seed-to-seed), not single-dataset posterior uncertainty.

The distance metrics above summarize point accuracy but not whether the posterior \emph{uncertainty} is well calibrated, which is one of the main advantages we claim for the joint Bayesian models over the two-step GG procedure. To assess this, we additionally report the empirical coverage of central 50\% and 90\% posterior credible intervals for the true turnout of each one- and two-way margin category (\autoref{sec:a2_coverage}). Because the main grid stored only per-chain posterior means, coverage was computed on a dedicated refit of the baseline scenario that retained the full posterior draws.

\subsection{Interval calibration}\label{sec:a2_coverage}
Using the baseline-scenario refit (all five models, 11 seeds, full posteriors retained), we computed the empirical coverage of central 50\% and 90\% posterior intervals for the true turnout of every one- and two-way margin category, pooled across categories and seeds (\autoref{tab:coverage}).

\begin{table}[ht]
\centering
\caption{Empirical coverage of central posterior credible intervals for true margin turnout, and mean 90\% interval width, pooled across categories and the 11 baseline datasets. Nominal coverage is 0.50 and 0.90; \emph{w90} is the mean width of the 90\% interval (smaller is more informative at equal coverage).}
\label{tab:coverage}
{\footnotesize
\begin{tabular}{lccccccc}
\toprule
 & \multicolumn{3}{c}{1D margins} & & \multicolumn{3}{c}{2D margins}\\
\cmidrule{2-4}\cmidrule{6-8}
Model & cov$_{50}$ & cov$_{90}$ & w90 & & cov$_{50}$ & cov$_{90}$ & w90\\
\midrule
BP & 0.01 & 0.03 & 0.133 & & 0.01 & 0.06 & 0.168\\
EI & 0.83 & 0.97 & 0.192 & & 0.61 & 0.93 & 0.539\\
GG & 0.74 & 0.93 & 0.187 & & 0.63 & 0.92 & 0.232\\
PM & 0.75 & 0.91 & 0.039 & & 0.35 & 0.77 & 0.097\\
FS & 0.78 & 0.96 & 0.039 & & 0.47 & 0.90 & 0.095\\
\bottomrule
\end{tabular}
}
\end{table}

Three patterns stand out. First, the survey-only BP model is grossly overconfident---its intervals are narrow but almost never contain the truth (coverage $\le 0.06$)---confirming that the survey bias it ignores is not reflected in its uncertainty. Second, EI and GG are close to nominal but pay for it with wide intervals (90\% widths of 0.19--0.54), reflecting the limited information in aggregate-only or two-step estimation. Third, and most importantly, PM and FS achieve comparable 90\% coverage with intervals 3--5 times narrower than GG (e.g.\ 1D width 0.039 vs.\ 0.187), so the joint models' lower point error is \emph{not} bought at the price of overconfident intervals; they are simply more informative. FS is also better calibrated than PM on the harder two-way margins (50\%/90\% coverage of 0.47/0.90 vs.\ 0.35/0.77), indicating that the selection correction tightens calibration where PM is mildly overconfident. Finally, FS reaches near-nominal 90\% coverage \emph{despite} its elevated $\hat{R}$, reinforcing that its mixing problems do not corrupt the turnout estimands.

\subsection{Scenarios and results}
This section defines the adjustment each scenario makes to the baseline data-generating process and presents its result. All scenarios share the baseline mechanism of the preceding subsections unless stated otherwise; the baseline parameters represent moderate selection bias and effect sizes: \(\sigma_{H,S}=\sigma_{H,O}=0.5\), \(\rho_{\varepsilon}=0.5\), \(N_S=1000\), \(\beta_0^s=-1\) (about 21\% selected), and \(\beta_0^o=0\) (about 50\% turnout). In every figure, points are medians and error bars the interquartile range across the 11 datasets.

\begin{figure}[H]
    \centering
    \caption{Baseline case: Kullback--Leibler divergence and total variation distance on a linear scale (the log-scale version is \autoref{fig:mb_default_parameter_metrics} in the main text). Points are medians and error bars the interquartile range across the 11 simulated datasets.}
    \label{fig:a2_default_parameter_metrics}
    \includegraphics[width=\textwidth]{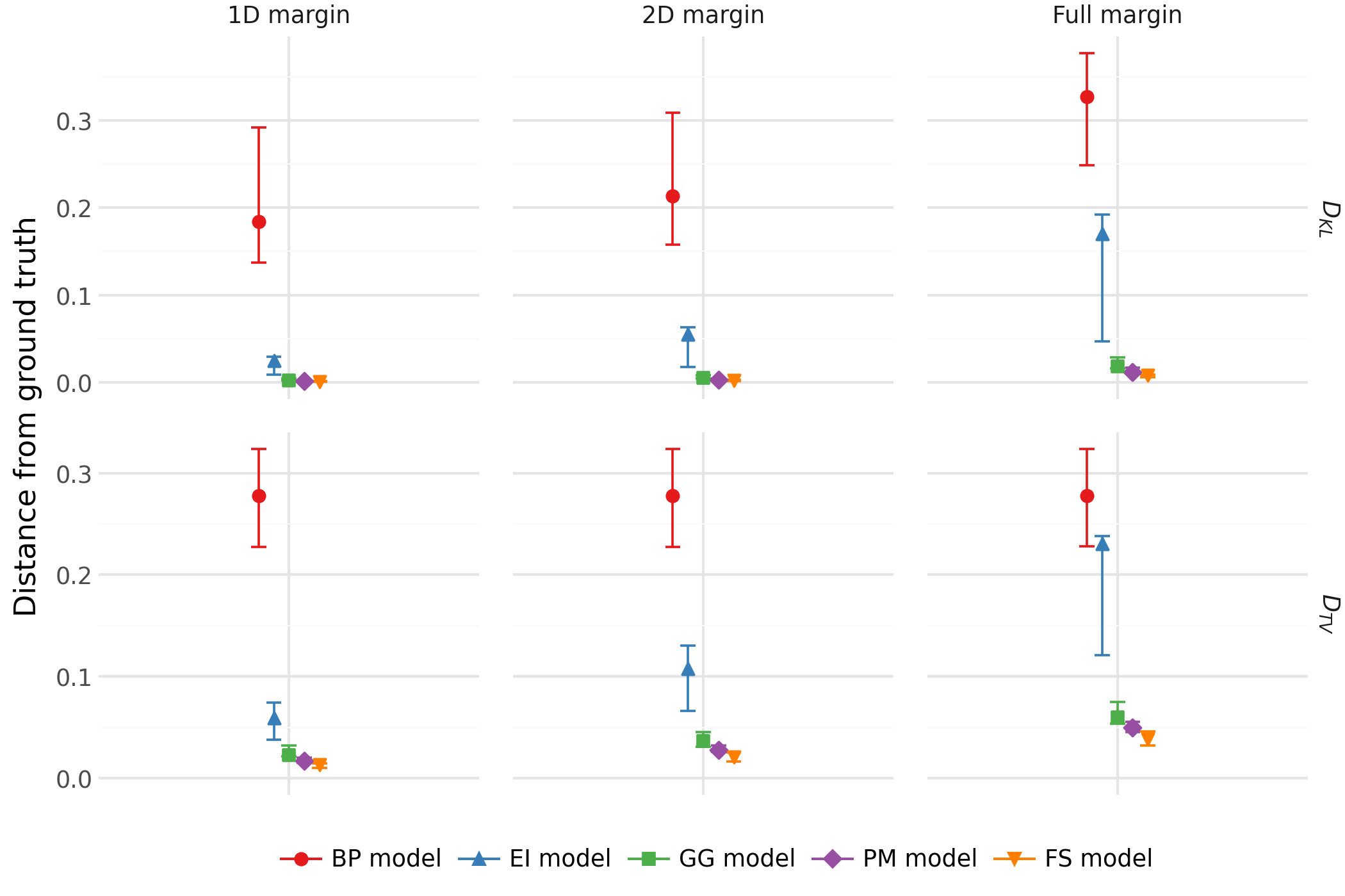}
\end{figure}

\Needspace*{0.58\textheight}
\paragraph{No-selection benchmark.}
We remove selection bias by omitting random effects in the selection equation (only an intercept), setting that intercept to zero, and setting the Heckman error correlation to zero ($\rho_{\varepsilon}=0$).

\begin{figure}[H]
    \centering
    \caption{With vs.\ without selection bias in the data-generating process.}
    \label{fig:a2_no_selection}
    \includegraphics[width=\textwidth]{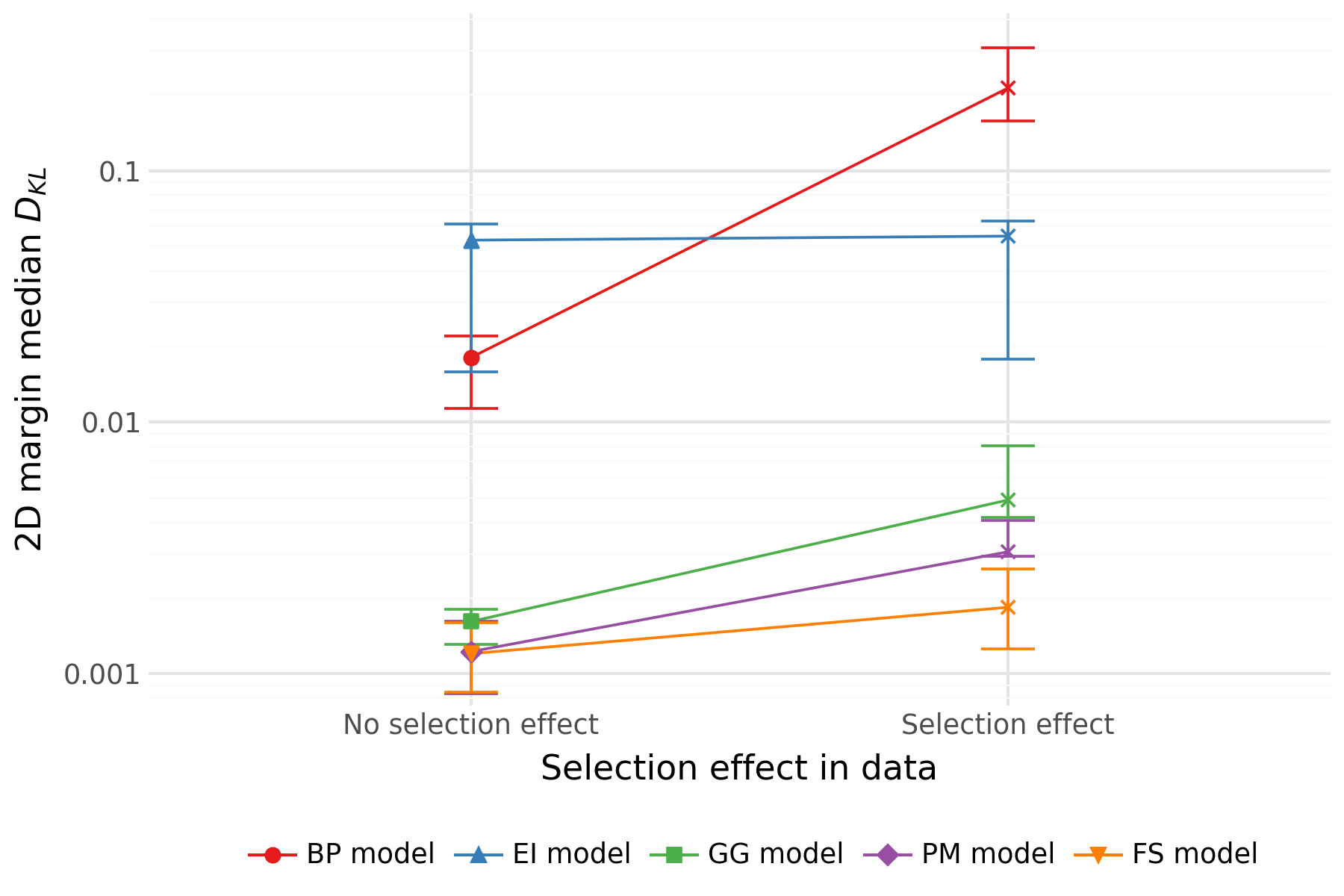}
\end{figure}

\paragraph{Varying censoring/response rate.}
We vary the selection intercept \(\beta_0^s\) over the grid \(\{-3,-2,\dots,3\}\), which changes the overall size of the selected subpopulation \(\{i:S_i^P=1\}\) from which the survey is drawn. In the write-up, \(\beta_0^s=-3\) corresponds to about 0.6\% selected, \(\beta_0^s=-1\) to about 21\%, \(\beta_0^s=0\) to about 50\%, and \(\beta_0^s=3\) to about 99\% (the figure reports realized selection rates averaged across seeds). The corresponding results are shown in \autoref{fig:mb_selection_bias_intercept_metrics} in the main text.

\Needspace*{0.58\textheight}
\paragraph{Error correlation.}
We vary the correlation \(\rho_{\varepsilon}\) in the bivariate Heckman error distribution over the grid \(\{0, 0.25, 0.5, 0.75, 1\}\), moving from independent selection and outcome errors (no selection bias) up to and including perfect correlation. All other baseline parameters are held fixed.

\begin{figure}[H]
    \centering
    \caption{Varying error correlation \(\rho_{\varepsilon}\) between selection and outcome.}
    \label{fig:a2_heck_cor}
    \includegraphics[width=\textwidth]{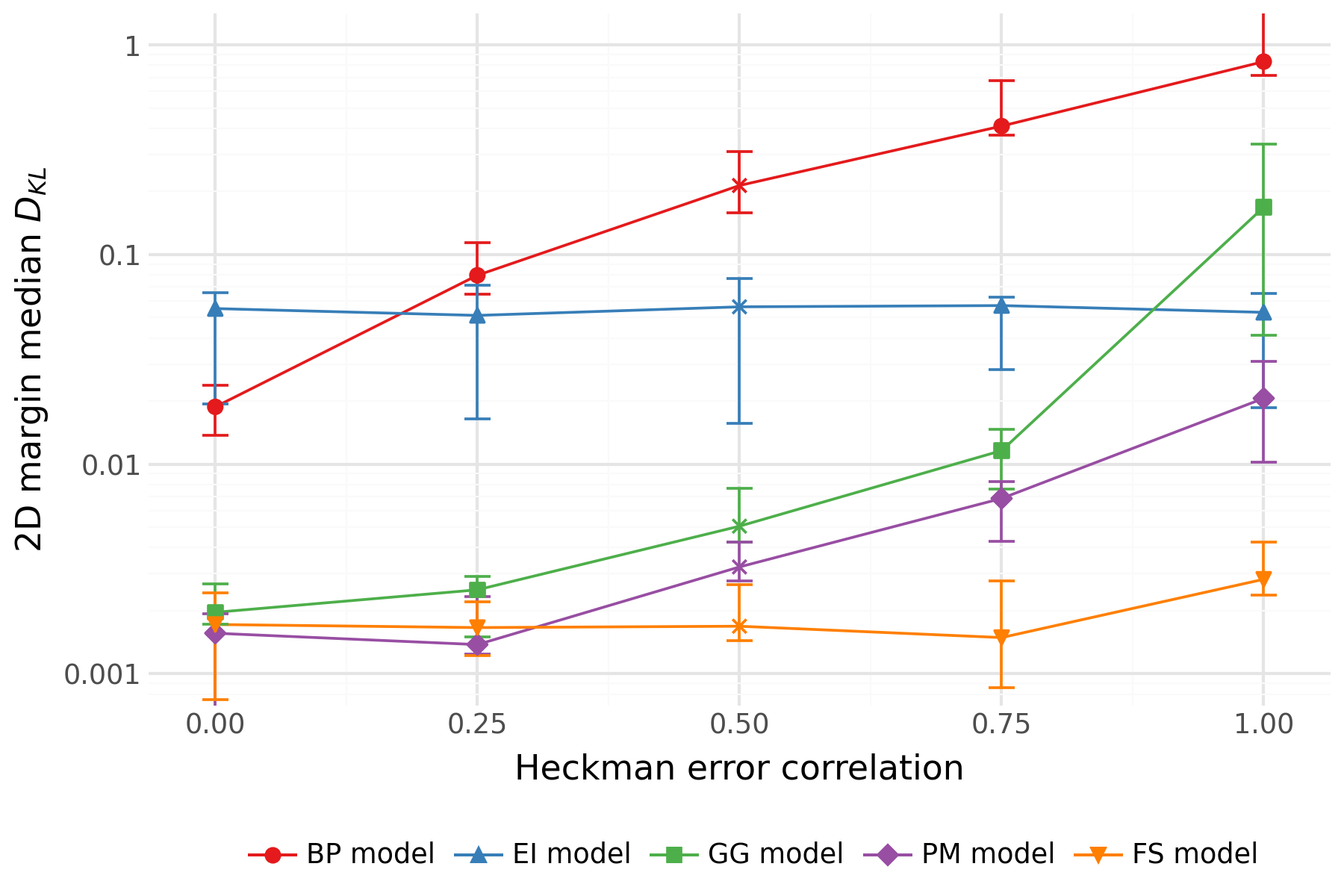}
\end{figure}

\Needspace*{0.58\textheight}
\paragraph{Measurement bias / over-reporting.}
We introduce over-reporting in survey responses by shifting only the survey reporting threshold. Concretely, we add a constant bias term \(\beta_{OB}\) to the latent outcome for selected individuals when generating the reported survey outcome \(O_i^S\), while leaving the true population outcome \(O_i^P\) unchanged (so \(O_i^P \neq O_i^S\) can occur). The simulation grid uses \(\beta_{OB}\in\{0,0.3,0.6,0.9\}\), which yields realized mis-reporting rates of about 0\%, 7\%, 14\%, and 19\% (averaged across seeds).

\begin{figure}[H]
    \centering
    \caption{Over-reporting / measurement bias in survey responses.}
    \label{fig:a2_overreport_bias}
    \includegraphics[width=0.9\textwidth]{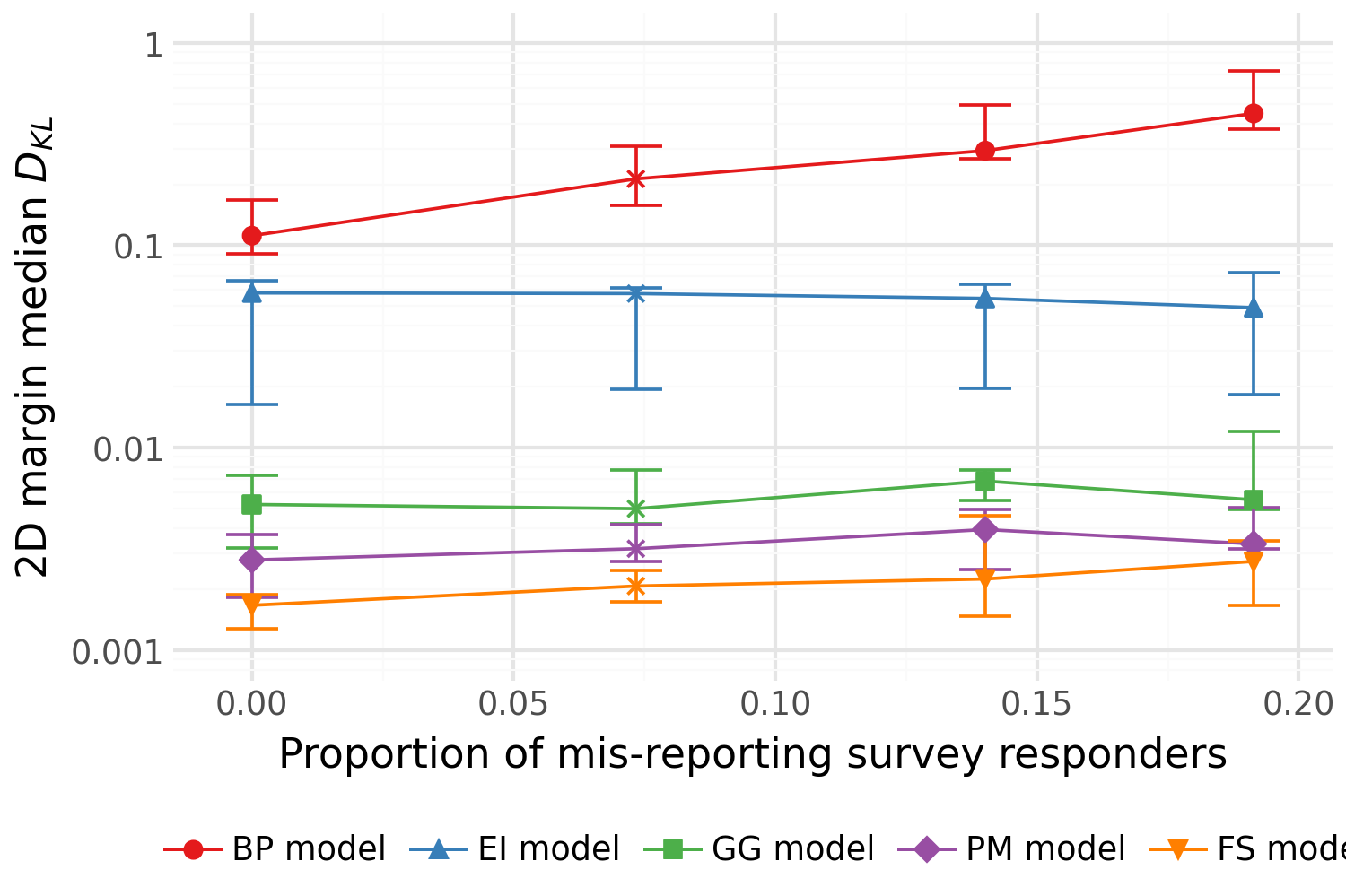}
\end{figure}

\Needspace*{0.58\textheight}
\paragraph{Aggregation bias.}
To simulate ecological aggregation bias (grouping-induced correlation between covariate margins and turnout), we augment the latent outcome for \emph{every} individual in region $g$ by the same amount, proportional to region $g$'s share of the nationality category ``Other'' (non-Estonian):
\[
O_i^* \gets O_i^* + \beta_{\mathrm{AB}}\, w_{g,\mathrm{Other}}\quad\text{for all } i \text{ with region } g,
\]
where $w_{g,\mathrm{Other}}$ is the proportion of ``Other'' in $g$ and $\beta_{\mathrm{AB}}$ is varied across simulation draws.

\begin{figure}[H]
    \centering
    \caption{Aggregation bias stress test.}
    \label{fig:a2_agg_bias}
    \includegraphics[width=\textwidth]{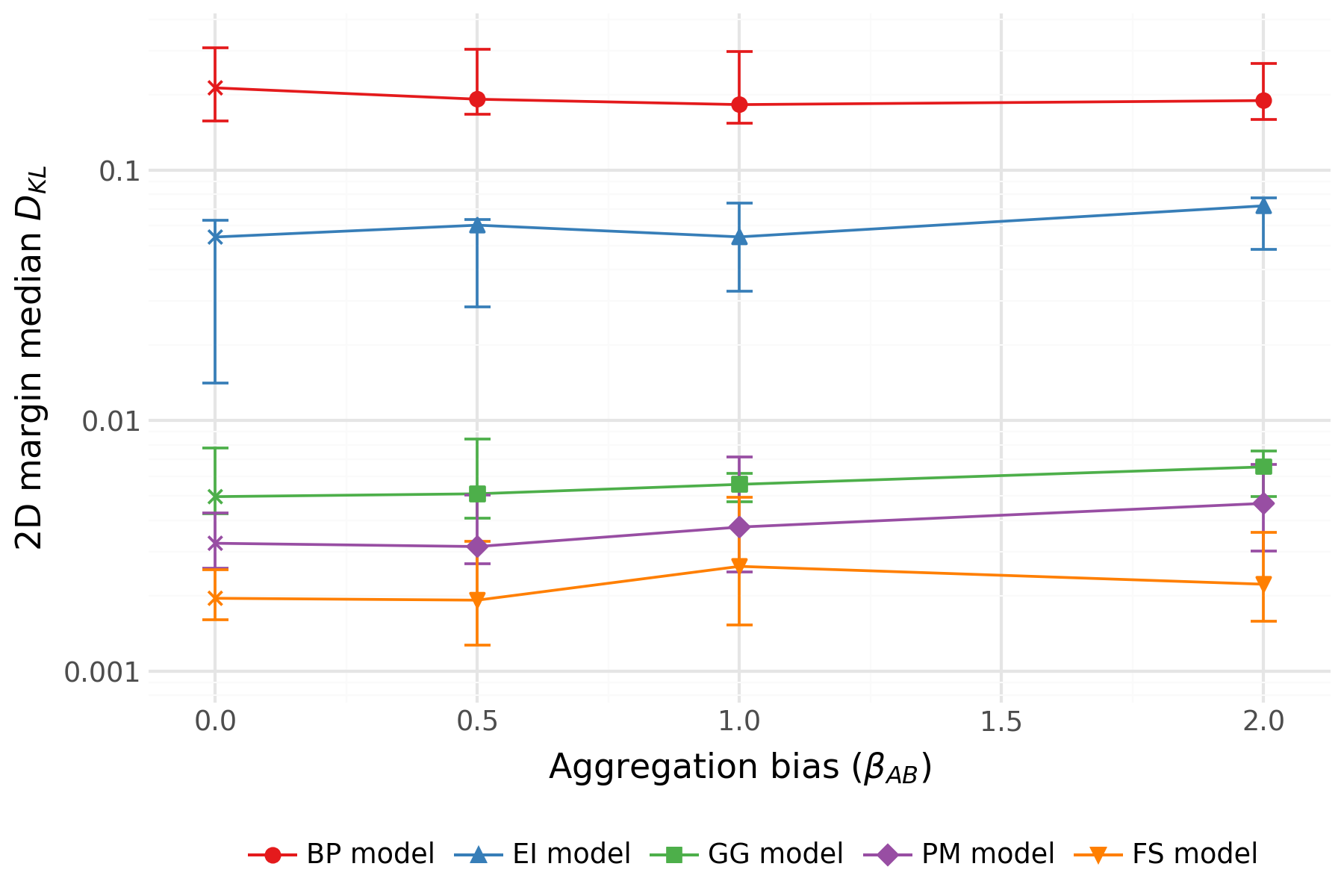}
\end{figure}

\Needspace*{0.68\textheight}
\paragraph{Collinearity between selection and outcome effects.}
To induce collinearity/identification difficulty, we draw the selection and outcome coefficients for each covariate-category pair jointly from a bivariate normal with correlation \(\rho_H\):
\[
\begin{pmatrix}\beta_{k,d}^s\\\beta_{k,d}^o\end{pmatrix} \sim \text{Normal} \left(
\begin{pmatrix}0\\0\end{pmatrix},
\begin{pmatrix}\sigma_{H,S}^2&\rho_H \sigma_{H,S} \sigma_{H,O}\\\rho_H \sigma_{H,S} \sigma_{H,O}&\sigma_{H,O}^2\end{pmatrix}
\right).
\]

\begin{figure}[H]
    \centering
    \caption{Correlated selection/outcome coefficients (collinearity stress test).}
    \label{fig:a2_hcoef_cor}
    \includegraphics[width=\textwidth]{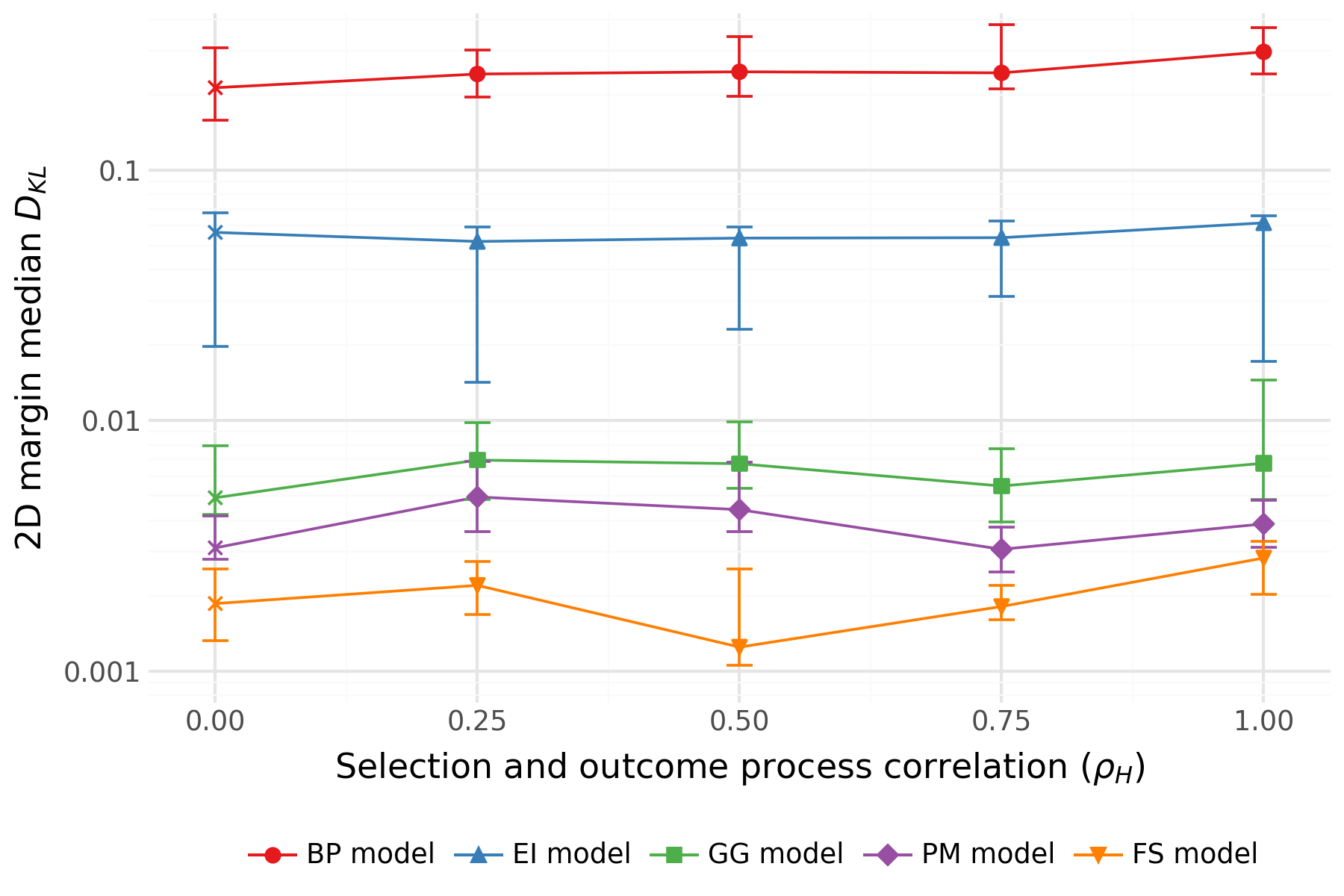}
\end{figure}

\Needspace*{0.68\textheight}
\paragraph{Non-normal errors: Skewness.}
We stress-test mis-specification of Gaussian selection/outcome errors using a bivariate skewed Student's $t$ with $\nu=3$ degrees of freedom, location $\mu$, dispersion $\Sigma$, and skewness vector $\gamma$, in the normal--variance mixture form used for the generalized hyperbolic skew-$t$ (Chapter~3 of McNeil, Frey and Embrechts, \textit{Quantitative Risk Management}, 2015):
\[
\begin{pmatrix}\varepsilon_i^s\\\varepsilon_i^o\end{pmatrix} = \mu + W_i \gamma + \sqrt{W_i}\, L z_i,
\qquad z_i \sim \mathcal{N}(0, I_2),
\qquad W_i \sim \mathrm{InvGamma}(\nu/2,\, \nu/2),
\]
independently of $z_i$, with $L$ the Cholesky factor of $\Sigma$ (so $\Sigma = L L^\top$). The symmetric Student-$t$ law is the special case $\gamma=\mathbf{0}$.
We do not include a separate pure (symmetric) Student-$t$ error scenario in the main replication grid; skewed-$t$ draws cover heavy tails and asymmetry jointly.

\begin{figure}[H]
    \centering
    \caption{Non-normal errors: skewed Student-\(t\) variants.}
    \label{fig:a2_non_normal_error}
    \includegraphics[width=\textwidth]{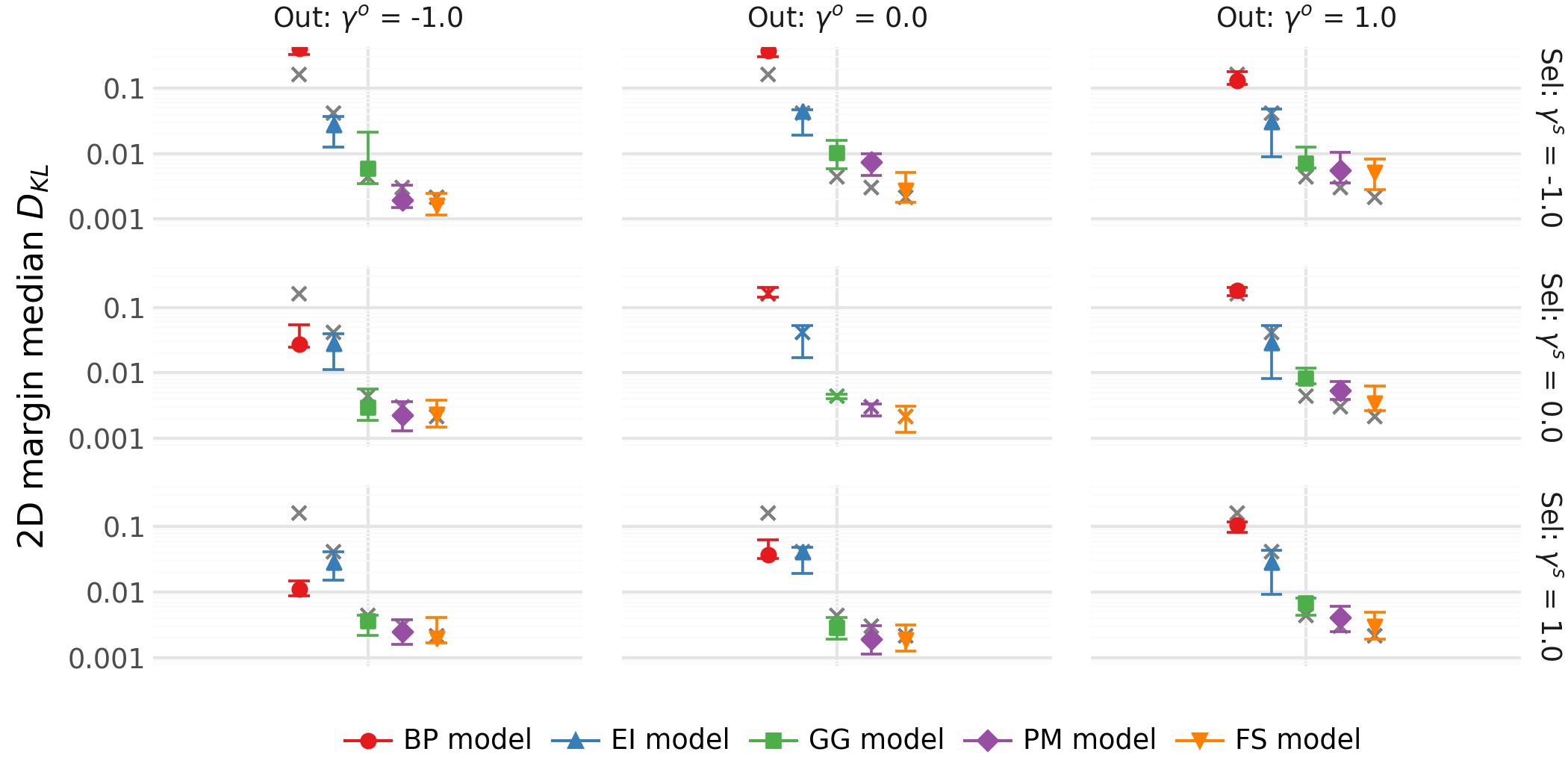}
\end{figure}

\begin{figure}[H]
    \centering
    \caption{Non-normal errors (skewed Student-$t$, $\nu=3$): alternative skew specifications.}
    \label{fig:a2_non_normal_error_tail}
    \includegraphics[width=\textwidth]{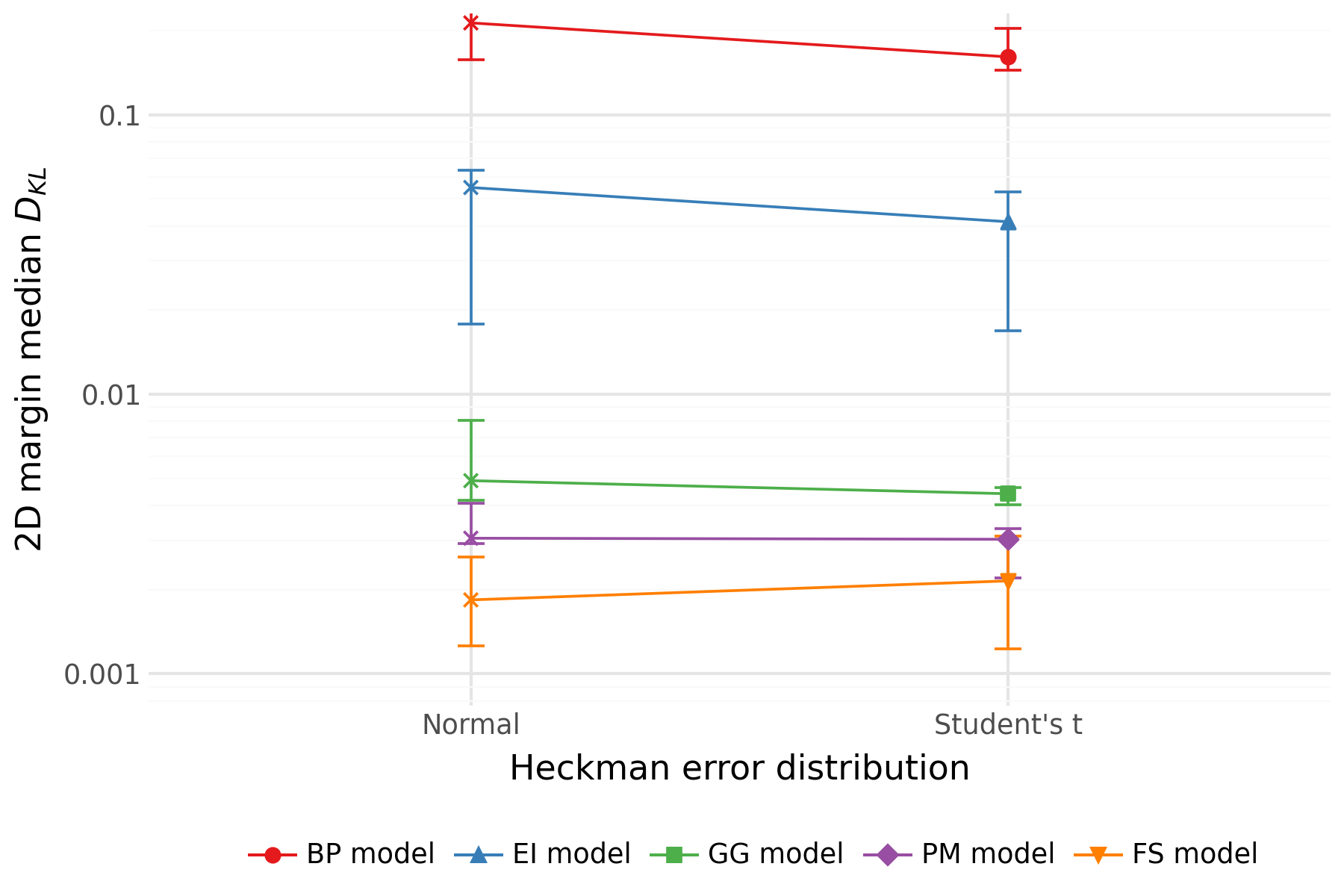}
\end{figure}

\Needspace*{0.58\textheight}
\paragraph{Selection and outcome effect sizes.}
We vary the typical magnitudes of selection vs.\ outcome effects by changing \(\sigma_{H,S}\) and \(\sigma_{H,O}\), which control the scale of the hierarchical coefficient distributions (through the half-normal hyperpriors on \(\tau_k^s\) and \(\tau_k^o\)).

\begin{figure}[H]
    \centering
    \caption{Varying selection and outcome effect sizes (\(\sigma_{H,S}, \sigma_{H,O}\)).}
    \label{fig:a2_hcoef_sigma_outcome}
    \includegraphics[width=\textwidth]{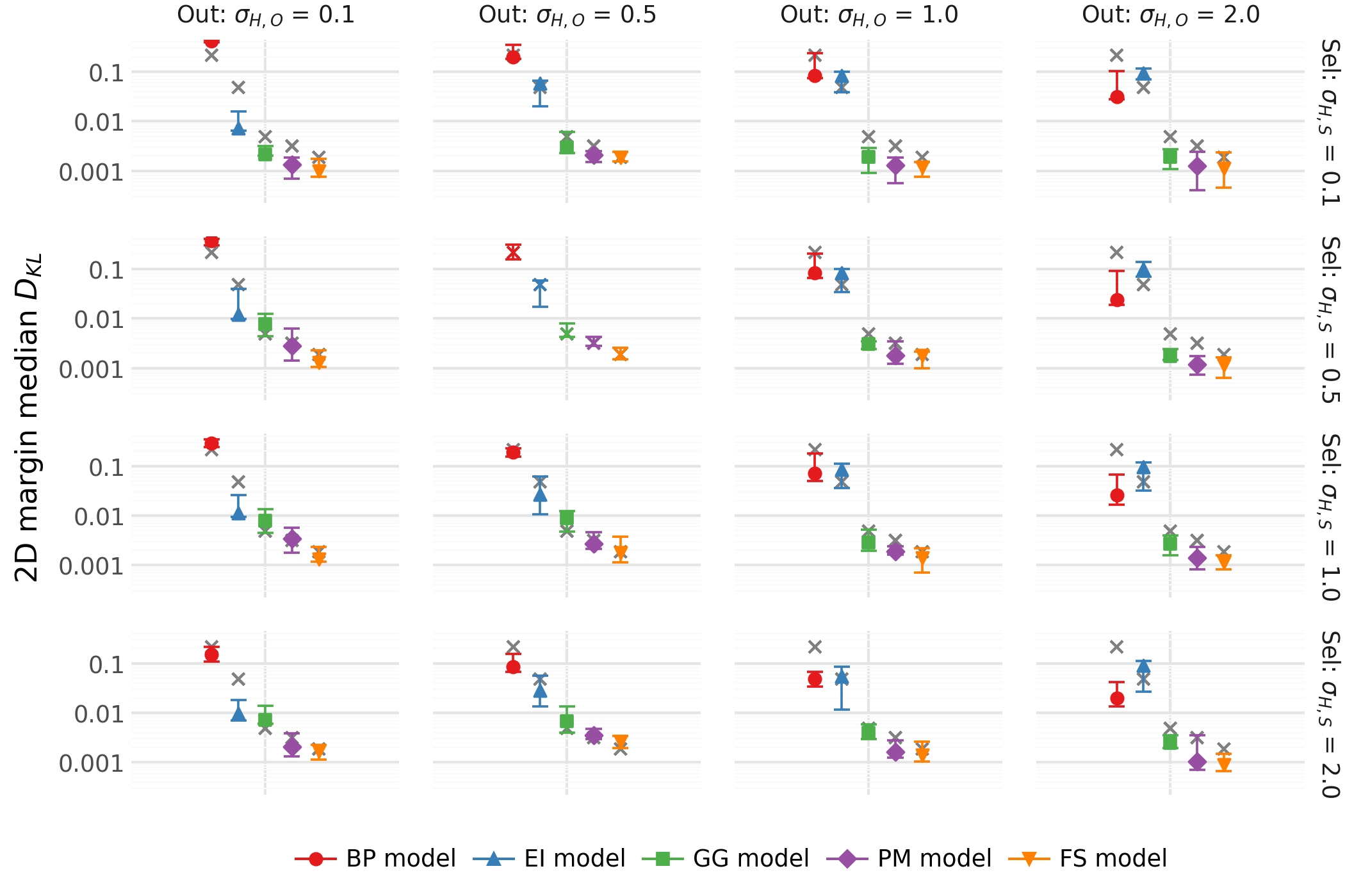}
\end{figure}

\Needspace*{0.58\textheight}
\paragraph{Prior informativeness (model adjustment).}
Separately from changing the data-generating process, we also compare weakly informative vs.\ more strongly informative priors in the fitted models (holding the data-generating process fixed), to test sensitivity to prior scale choices.

\begin{figure}[H]
    \centering
    \caption{Prior informativeness sensitivity.}
    \label{fig:a2_prior_scale}
    \includegraphics[width=0.9\textwidth]{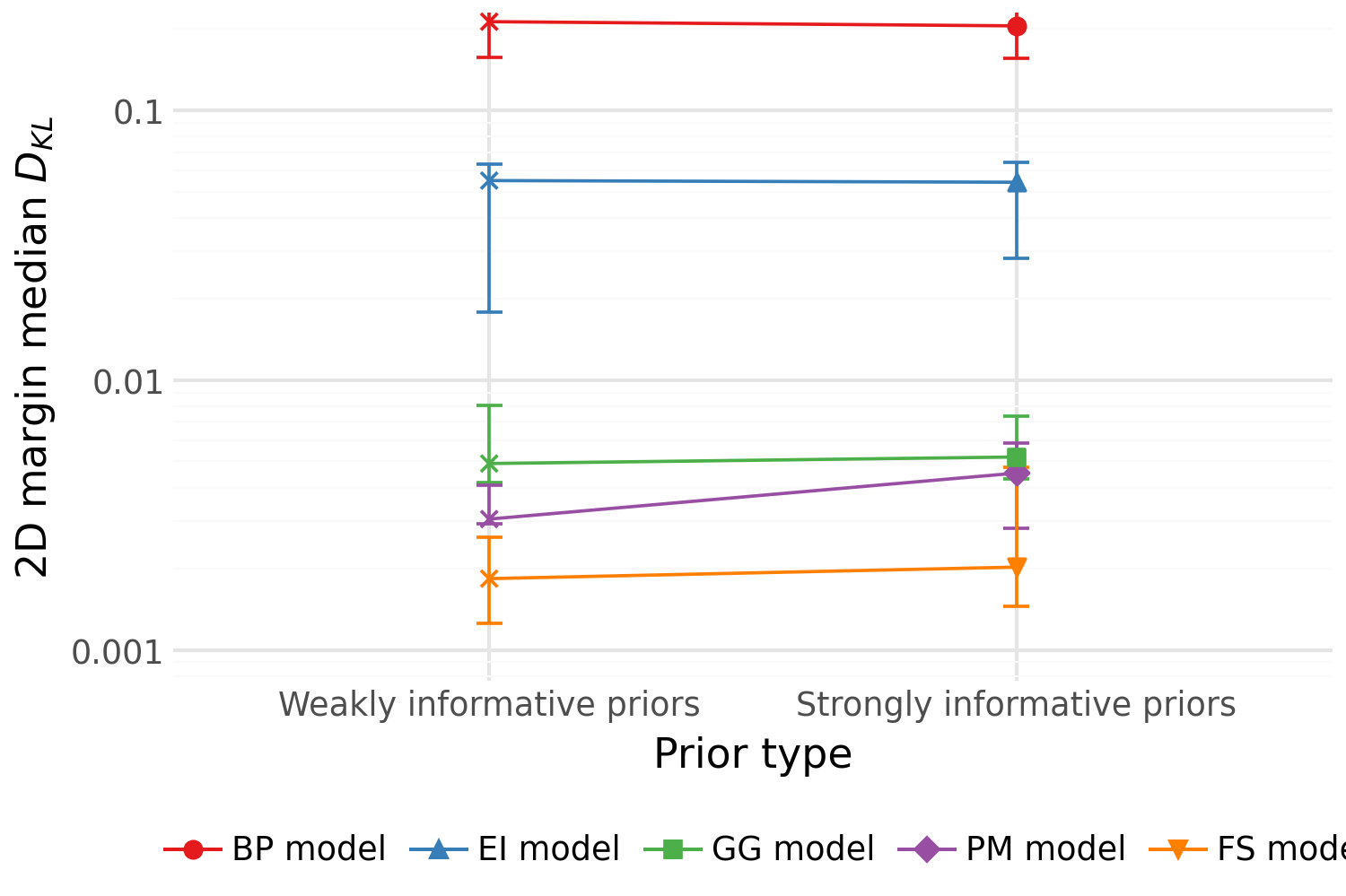}
\end{figure}

\Needspace*{0.68\textheight}
\paragraph{Interaction effects.}
We introduce interaction effects by adding pairwise interaction coefficients to both selection and outcome linear predictors:
\begin{align*}
S_i^* &= X_i \beta^s + \sum_{k,l \in \mathcal{I}} \beta_{k,l}^s x_k x_l + \varepsilon_i^s\\
O_i^* &= X_i \beta^o + \sum_{k,l \in \mathcal{I}} \beta_{k,l}^o x_k x_l + \varepsilon_i^o,
\end{align*}
where \(\mathcal{I}\) contains all pairwise interactions among the five demographic inputs (age, gender, education, nationality, region).

\begin{figure}[H]
    \centering
    \caption{Interaction effects: with vs.\ without pairwise interactions.}
    \label{fig:a2_int}
    \includegraphics[width=0.9\textwidth]{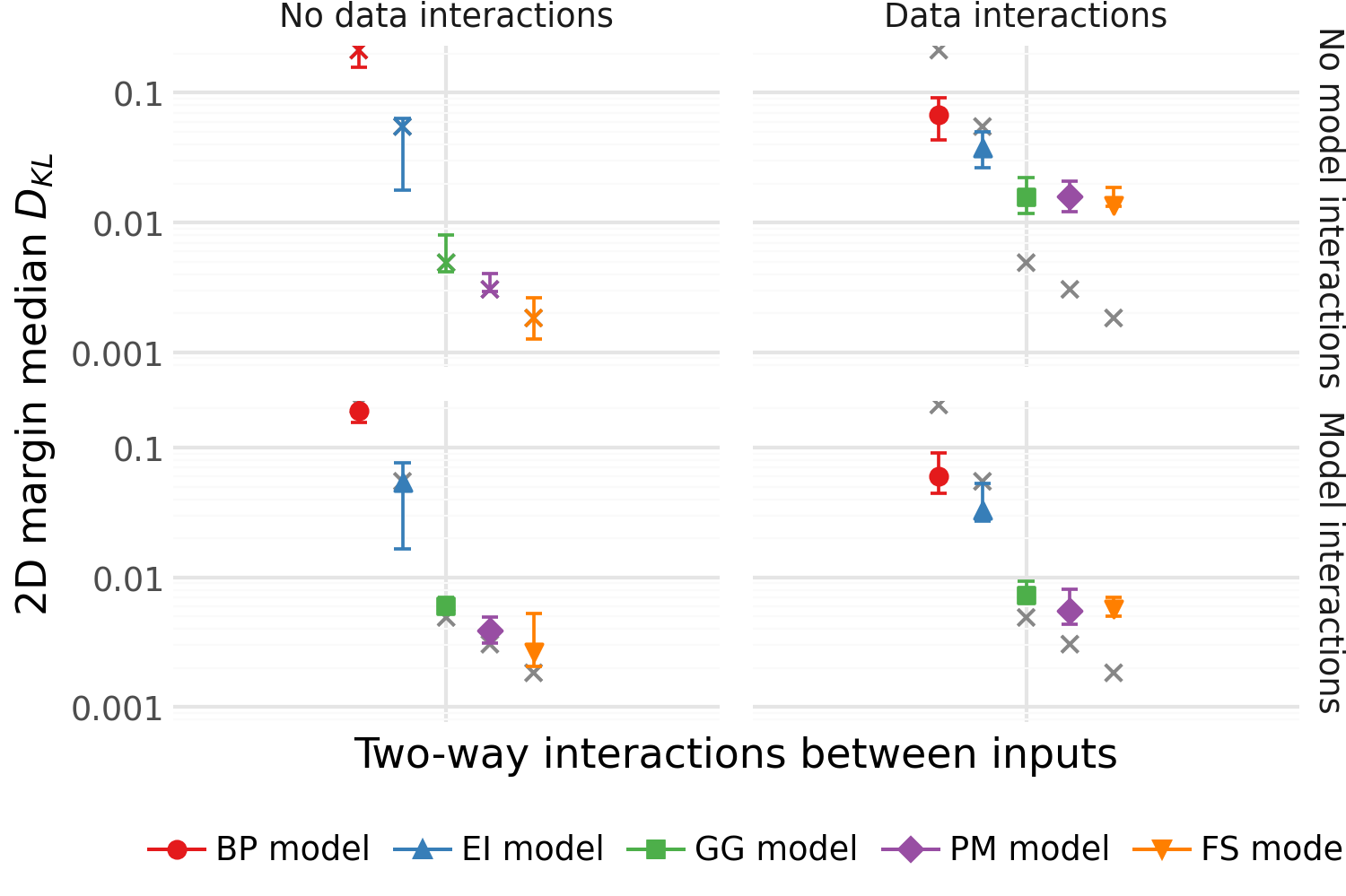}
\end{figure}

\Needspace*{0.58\textheight}
\paragraph{Random noise.}
For selection, we implement an asymmetric perturbation of the probit selection rule: among individuals who would not select into the survey under the latent index, a random fraction may nevertheless be flipped to $S_i^P=1$, with the flip probability calibrated so the overall contamination rate matches the nominal scenario. For outcomes, we perturb a random subset of units by Bernoulli noise on $O_i^P$. The two noise types are varied independently, each with the other held at zero (the simulation grid does not cross them).

\begin{figure}[H]
    \centering
    \caption{Random noise in the selection mechanism (left) and the outcome mechanism (right).}
    \label{fig:a2_noise_outcome}
    \includegraphics[width=\textwidth]{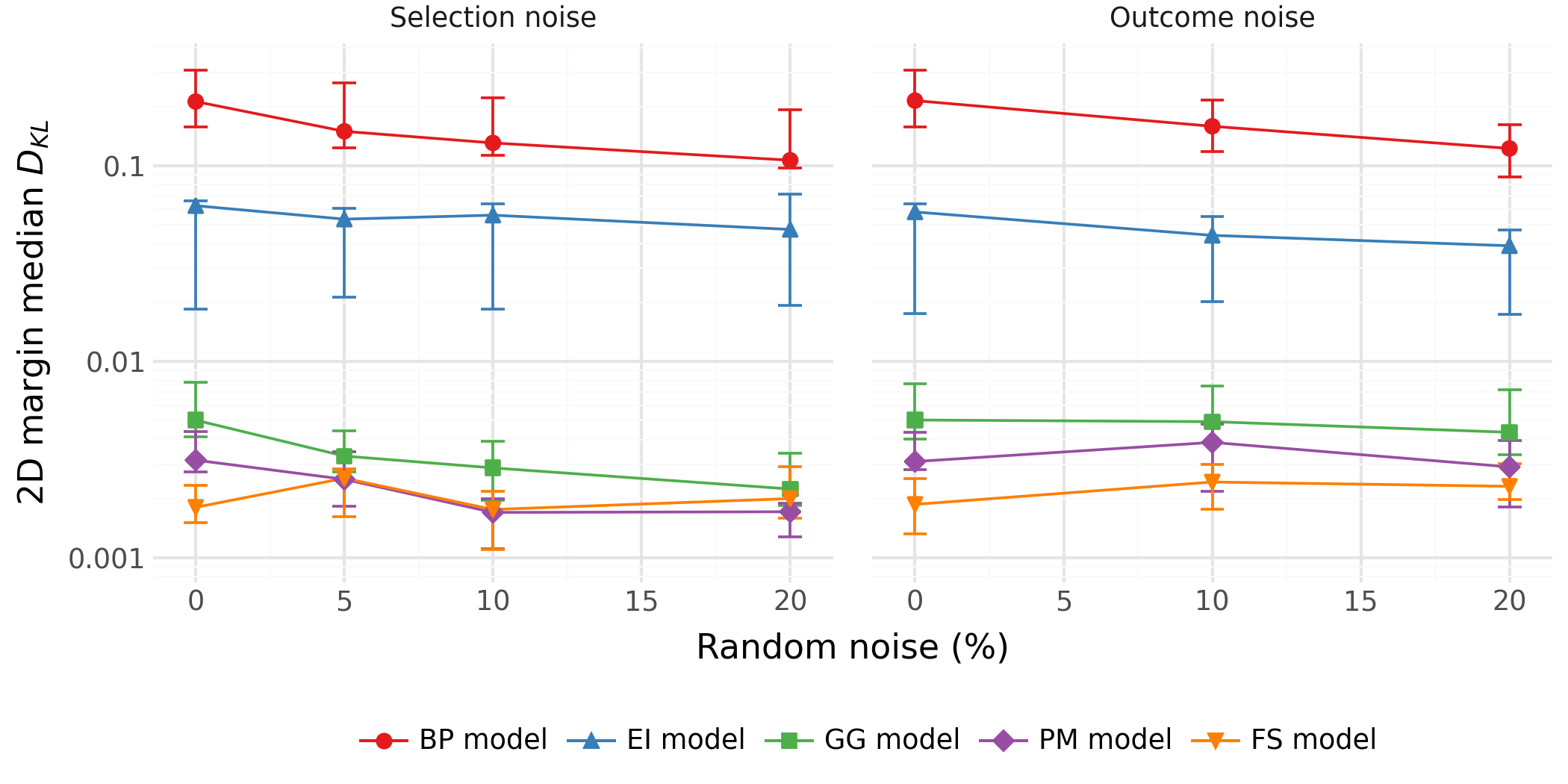}
\end{figure}

\Needspace*{0.58\textheight}
\paragraph{Sample size.}
We vary the survey sample size \(N_S\), while keeping the selection mechanism (and thus the selected subpopulation definition) fixed, and still drawing the survey uniformly from \(\{i:S_i^P=1\}\).

\begin{figure}[H]
    \centering
    \caption{Varying survey sample size \(N_S\).}
    \label{fig:a2_sample_size}
    \includegraphics[width=\textwidth]{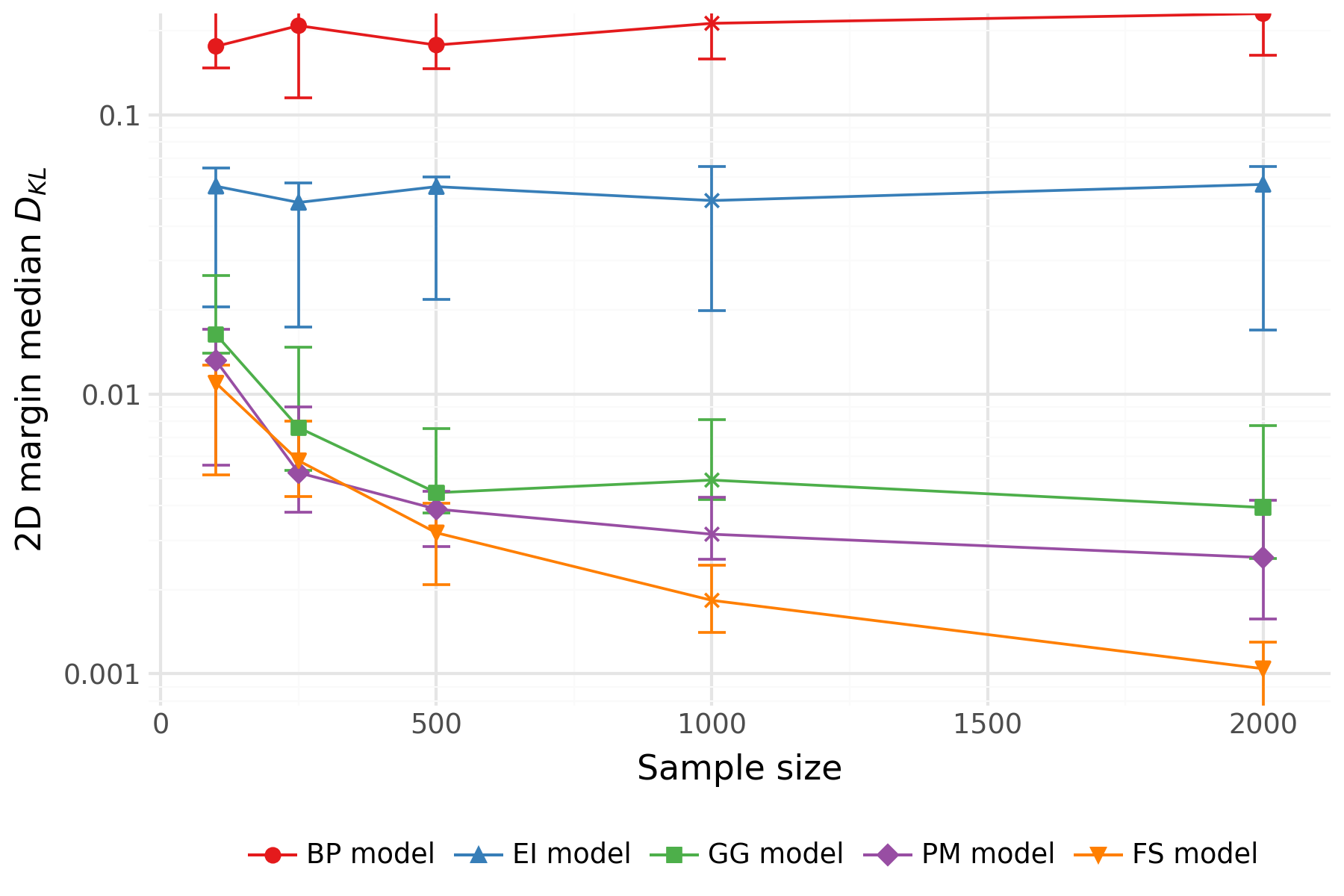}
\end{figure}

\Needspace*{0.58\textheight}
\paragraph{Margin informativeness.}
We vary which aggregate turnout margins are provided to the models, from least to most informative. The conditions are (i) topline-only turnout, (ii) electoral-district margins, and (iii) region margins. Unlike the other scenarios, the data for this scenario is generated with electoral district included as an additional covariate in both the selection and outcome processes, so that district-level margins carry information about the data-generating process.

\begin{figure}[H]
    \centering
    \caption{Varying aggregate margin informativeness.}
    \label{fig:a2_margin}
    \includegraphics[width=0.9\textwidth]{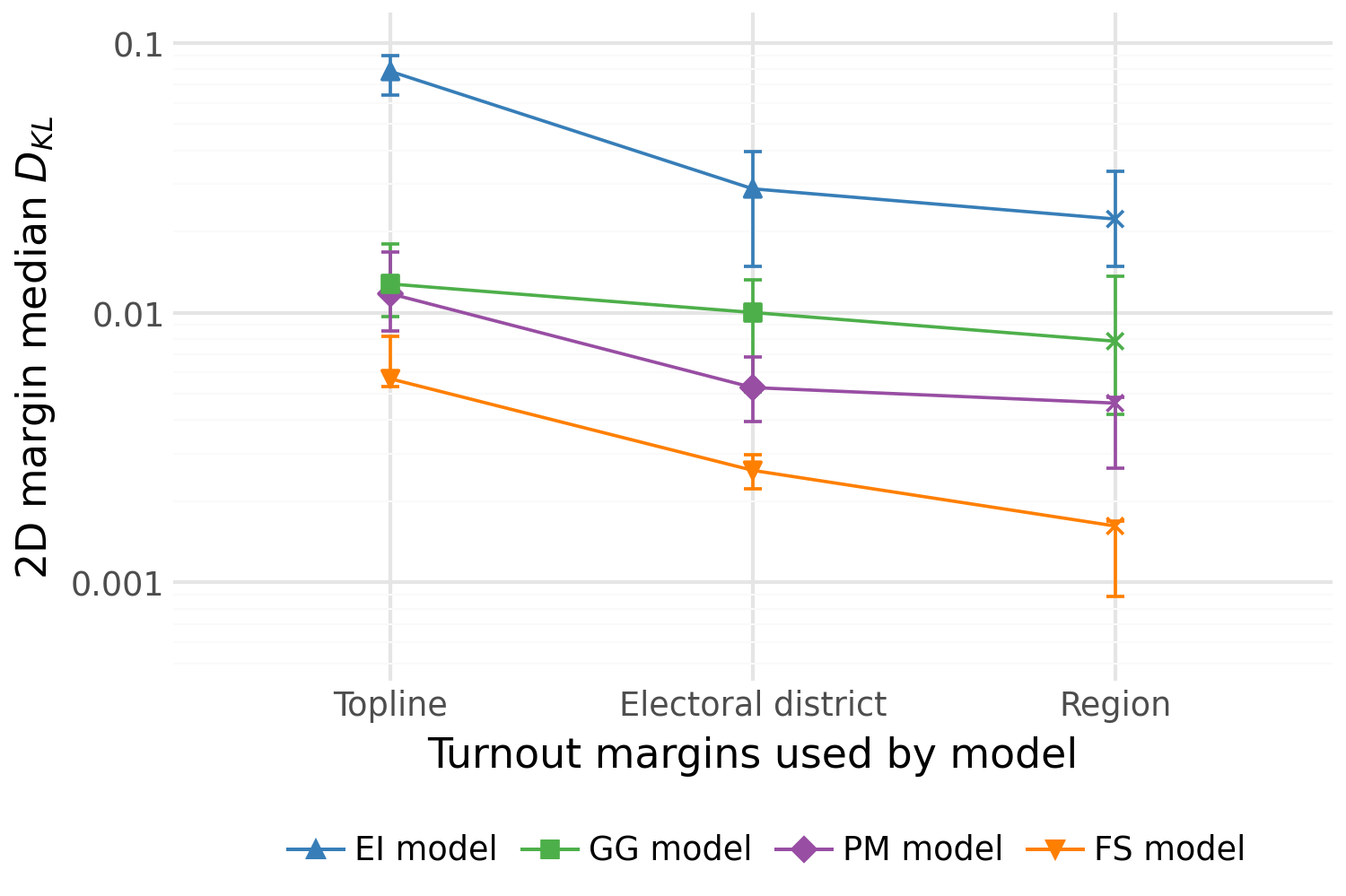}
\end{figure}

\subsection{Erratum: placement of the overreporting term in FS}
The FS implementation used for the simulations and the empirical application applied the overreporting shift \(\beta^o_{or}\) to the \emph{conditional} outcome probability, computing \(\Phi\big((\eta^o_i + \rho u_i)/\sqrt{1-\rho^2} + \beta^o_{or}\big)\) rather than the specified \(\Phi\big((\eta^o_i + \beta^o_{or} + \rho u_i)/\sqrt{1-\rho^2}\big)\) (since corrected in the released package). This cannot materially affect the results: for any fixed \(\rho\) the two forms are related by the one-to-one rescaling \(\beta^o_{or} \mapsto \beta^o_{or}\sqrt{1-\rho^2}\), so they define identical families of conditional outcome probabilities, and the turnout estimands depend only on \(\Phi(\eta^o_c)\), never on \(\beta^o_{or}\) directly.

\section{Model extensions} \label{sec:appendix_extensions}

This appendix details the extensions implemented in the companion package and summarized in the main text.

\paragraph{Ordinal outcomes.}
While the main text focuses on binary voting intention, surveys often use Likert-scale questions with \(K\) ordered categories to capture voting likelihood. Our framework can accommodate this by replacing the binary probit model for \(O_i\) with a generalized ordinal probit model. This requires introducing cutpoints \(\kappa_k \sim \mathcal{N}(0,\sigma_c)\) subject to the ordering constraint \(\kappa_1<\cdots<\kappa_{K-1}\). The outcome probabilities are then given by \(P(O_i=k \mid X_i)=\Phi(\kappa_k-\eta_i^o)-\Phi(\kappa_{k-1}-\eta_i^o)\) for \(k=1,\dots,K\), with \(\kappa_0=-\infty\) and \(\kappa_K=\infty\). In the binary case, this conveniently reduces to the original model with \(\kappa_1 \equiv -\beta^o_{or}\).

\paragraph{Multiple polls and margins.}
The framework naturally supports the integration of multiple polls and diverse aggregate margins (e.g., age, gender, and region simultaneously). This is achieved by repeating the outcome likelihoods---Equation \eqref{eq:bernoulli_obs} for polls and Equation \eqref{eq:margin_obs} for margins (or their FS equivalents)---for each additional data source. Although this approach technically assumes independence between polls and margins, which may be a simplification, it serves as a practical approximation for leveraging all available data.

\paragraph{Beta-binomial margin likelihood.}
The binomial likelihood \eqref{eq:margin_obs} treats the official counts as exact binomial draws from the model-implied turnout probabilities. With regional populations in the tens of thousands, the implied relative uncertainty is of order \(1/\sqrt{N_r}\), so the margins act as near-hard constraints on the poststratified estimates. This is appropriate when the census cell counts and the turnout denominators refer to exactly the same population, but administrative counts are not random draws, and in practice the two frames rarely match perfectly (e.g., voters living abroad, register lag, or voting outside one's home district, as in our application). Any such frame mismatch is then forced into the outcome coefficients rather than absorbed as noise. To make the margin fit more forgiving, the margin likelihood can be replaced with a beta-binomial with a learned concentration parameter; we report a preliminary evaluation on the Estonian data in \autoref{sec:application}, but a systematic evaluation of this variant across scenarios is left for future work.

\paragraph{Robust selection.}
Following the promising results of \citet{MarchenkoGenton} and \citet{ding2014}, the selection mechanism can employ a Student's \(t\)-distribution with a learned degrees of freedom parameter \(\nu\) instead of a normal distribution. While this modification significantly increases computational cost (and was therefore excluded from our simulation studies), it offers a pathway to better handle outliers and noise in the selection process.

\end{document}